\documentclass[a4paper,fleqn,usenatbib]{mnras}
\usepackage{newtxtext,newtxmath}
\usepackage[T1]{fontenc}
\usepackage{ae,aecompl}
\usepackage{graphicx}	
\usepackage{amsmath}	

\usepackage{amssymb}	
\usepackage{mathrsfs}
\usepackage{longtable}
\usepackage{hyperref}
\usepackage[font=small]{caption}
\usepackage{bm}

\title[SN~2019yvr]{An environmental analysis of the Type~Ib SN~2019yvr and the possible presence of an inflated binary companion}

\author[N.-C. Sun et al.]{Ning-Chen Sun$^1$\thanks{E-mail: n.sun@sheffield.ac.uk}, Justyn R. Maund$^1$, Paul A. Crowther$^1$, Ryosuke Hirai$^{2, 3}$,
\newauthor
Amir Kashapov$^3$, 
Ji-Feng Liu$^4$, Liang-Duan Liu$^5$, and Emmanouil Zapartas$^6$ \\
1 Department of Physics and Astronomy, University of Sheffield, Hicks Building, Hounsfield Road, Sheffield S3 7RH, UK \\
2 OzGrav: Australian Research Council Centre of Excellence for Gravitational Wave Discovery, Clayton, VIC 3800, Australia \\
3 School of Physics and Astronomy, Monash University, Clayton, Victoria 3800, Australia \\
4 National Astronomical Observatories, Chinese Academy of Sciences, 20A Datun Road, Chaoyang District, Beijing 100101, China \\
5 College of Physical Science and Technology, Central China Normal University, 152 Luoyu Road, Wuhan, Hubei 430079, China \\
6 D\'epartement d'Astronomie, Université de Gen\`eve, Chemin Pegasi 51, CH-1290 Versoix, Switzerland
}

\date{Accepted XXX. Received YYY; in original form ZZZ}

\pubyear{2020}

\begin{document}
\label{firstpage}
\pagerange{\pageref{firstpage}--\pageref{lastpage}}
\maketitle

\begin{abstract}

SN~2019yvr is the second Type~Ib supernova (SN) with a possible direct detection of its progenitor (system); however, the spectral energy distribution (SED) of the pre-explosion source appears much cooler and overluminous than an expected helium-star progenitor. Using Hubble Space Telescope (HST) images and MUSE integral-field-unit (IFU) spectroscopy, we find the SN environment contains three episodes of star formation; the low ejecta mass suggests the SN progenitor is most likely from the oldest population, corresponding to an initial mass of 10.4$^{+1.5}_{-1.3}$~$M_\odot$. The pre-explosion SED can be reproduced by two components, one for the hot and compact SN progenitor and one for a cool and inflated yellow hypergiant (YHG) companion that dominates the brightness. Thus, SN~2019yvr could possibly be the first Type~Ib/c SN for which the progenitor's binary companion is directly detected on pre-explosion images. Both the low progenitor mass and the YHG companion suggest significant binary interaction during their evolution. Similar to SN~2014C, SN~2019yvr exhibits a metamorphosis from Type~Ib to Type~IIn, showing signatures of interaction with hydrogen-rich circumstellar material (CSM) at $>$150~days; our result supports enhanced pre-SN mass loss as an important process for hydrogen-poor stars at the lower-mass end of core-collapse SN progenitors.

\end{abstract}


\defcitealias{K21}{K21}
\defcitealias{Yoon2017}{Y17}

\section{Introduction}
\label{intro.sec}

Core-collapse SNe are spectacular explosions of massive ($>$8~$M_\odot$) stars at the end of their lives; understanding the progenitors of different SN types is of vital importance for the research of massive stars' evolution and final explosions. While the progenitors of the most common Type~II-P SNe have been (relatively) well studied \citep{Smartt2009}, the direct detection of Type~Ib/c SN progenitors has been extremely challenging. The lack of hydrogen (and even helium) features in Type~Ib/c spectra suggests that their progenitors' outer envelopes have been stripped; as a result, their progenitors are much hotter and more compact than the red supergiant (RSG) progenitors for Type~II-P SNe, and thus much more difficult to be detected in the optical filters. Type~Ib/c SNe are also found in closer vicinity to star-forming regions where the high extinctions from the dusty interstellar medium may obscure the light of their progenitors \citep{Maund2018}. Until now, there are only three Type~Ib/c SNe with (possible) progenitor detections, i.e. the Type~Ib iPTF13bvn \citep{Cao2013, Eldridge2015, Eldridge2016, Folatelli2016}, Type~Ic SN~2017ein \citep{Kilpatrick2018, VanDyk2018, Xiang2019}, and Type~Ib SN~2019yvr (\citealt{K21}; \citetalias{K21} hereafter) that will be discussed in this paper.

Theoretically, there could be two possible progenitor channels to produce a stripped-envelope SN. A star's envelope can be ejected via its own stellar wind if it is massive enough ($>$~25~$M_\odot$ at solar metallicity) to evolve to a classical Wolf-Rayet star \citep{Gaskell1986, Crowther2007}. Alternatively, a moderately massive progenitor can be stripped in interacting binaries via Roche-lobe overflow, common-envelope evolution \citep{Pods1992}, or even by the impact of the companion star's SN explosion \citep{Hirai2020}. Statistical analysis of the occurrence rate \citep{Smith2011, Graur2017, Shivvers2017} and ejecta mass \citep{Drout2011, Taddia2015, Taddia2018, Lyman2016, Prentice2016, Prentice2019} suggests that the binary progenitor channel plays a significant, and maybe even a dominant role for Type~Ib/c SNe.

This conclusion motivates astronomers to search for the binary companions of nearby SNe. Currently, there are only four core-collapse SNe with direct companion detections, i.e. the Type~IIb SN~1993J \citep{Maund2004, Fox2014}, SN~2001ig \citep{Ryder2018} and SN~2011dh \citep{Folatelli2014, Maund2015, Maund2019}, and the Type~Ibn SN~2006jc \citep{Maund2016b, Sun2020a}. The detected companions can place tight constraints on the initial properties and the pre-SN evolution of the SN progenitor systems. For example, the companion of SN~1993J observationally confirmed an interacting binary progenitor model for this Type~IIb SN \citep{Maund2004}, which was theoretically proposed soon after its discovery \citep{Nomoto1993, Pods1993}. \citet{Sun2020a} derived an upper age limit for SN~2006jc's companion, which favours a moderately massive binary progenitor system and argues against the prevailing theory of very massive WR progenitors for Type~Ibn SNe. For Type~Ib/c SNe, however, there have been no companion detections reported yet.

SN~2019yvr is a Type~Ib SN in the nearby galaxy NGC~4666. At $\sim$2.6~years before its explosion, \citet{Graur2018} acquired deep and high-spatial resolution images of NGC~4666 with the HST at five epochs and in four filters in order to study the late-time brightness of the Type~Ia SN ASASSN-14lp that also occurred in this galaxy. With these images, \citetalias{K21} detected the progenitor (system) candidate of SN~2019yvr with high signal-to-noise ratios (SNRs). Very interestingly, however, the SED of this source appears much cooler and more luminous than a hot and compact Type~Ib SN progenitor. Thus, there is a significant tension between SN~2019yvr's progenitor detection and the current theories of Type~Ib SNe.

Despite the hydrogen-poor nature of SN~2019yvr at early times, from $>$150~days post discovery it exhibited strong, narrow H$\alpha$ emission resulting from the interaction with hydrogen-rich CSM ejected only decades before core collapse (\citetalias{K21}). This makes SN~2019yvr a member of the rare SN~2014C-like class with spectral metamorphosis from Type~Ib to Type~IIn \citep{Mili2015, Margutti2017}. While a handful of such SNe have been discovered \citep[e.g. SN~2004dk,][]{Mauerhan2018, Pooley2019}, only for SN~2014C has the progenitor been relatively well constrained \citep[by analysing its host star cluster,][]{Sun2020b}. Thus, SN~2019yvr provides another precious opportunity to investigate the progenitors of these peculiar SNe.

Environmental analysis serves as a powerful tool to study the progenitors of core-collapse SNe (e.g. \citealt{Maund2018, Sun2021}). Since most massive stars form in groups, we expect the progenitor to have the same age as the nearby stars that have been born together in the same burst of star formation. In this paper, we carry out a detailed investigation of SN~2019yvr's environment using a combination of high-spatial resolution images and spatially resolved IFU spectroscopy. Our aim is to constain its progenitor properties and resolve the mystery of its pre-explosion source.

Throughout this paper, we adopt a distance modulus of 30.8~$\pm$~0.2~mag \citep[i.e. 14.4~$\pm$~1.3~Mpc;][]{distance.ref}, a redshift of $z$~= 0.005080 \citep{redshift.ref}, and a Galactic foreground reddening of $E(B-V)$~= 0.02~mag \citep{galebv.ref} for SN~2019yvr's host galaxy, all consistent with \citetalias{K21}. SN~2019yvr's extinction within the host galaxy is large and \citetalias{K21} derived $A_V$~=~2.4$^{+0.7}_{-1.1}$ for a total-to-selective extinction of $R_V$~=~4.7$^{+1.3}_{-3.0}$.

This paper is structured as follows. Section~\ref{data.sec} describes the data used and the SN environment is explored in Section~\ref{env.sec}. We investigate the bolometric light curve in Section~\ref{curve.sec} and analyse the pre-explosion progenitor detection in Section~\ref{detection.sec}. The results are followed by a discussion in Section~\ref{discussion.sec} before we close the paper with a summary and conclusions in Section~\ref{summary.sec}.

\section{Data}
\label{data.sec}

\begin{table}
\centering
\caption{HST/WFC3/UVIS observations of the site of SN~2019yvr before its explosion (program ID: 14611; PI: Graur~O.).}
\begin{tabular}{cccc}
\hline
\hline
Date & Time from $V$-band & Filter & Exposure \\
(UT) & peak brightness (day)  & & Time (s) \\
\hline
2018-04-21 & $-$989
& F438W &  1140 \\
& & F555W &  1200 \\
& & F625W &  1134 \\
& & F814W &  1152 \\
\hline
2018-05-17 & $-$963
& F555W &  1143 \\
& & F625W &  1140 \\
\hline
2018-06-13 & $-$936
& F438W &  1140 \\
& & F555W &  1200 \\
& & F625W &  1134 \\
& & F814W &  1152 \\
\hline
2018-07-10 & $-$908
& F555W &  1143 \\
& & F625W &  1140 \\
\hline
2018-08-07 & $-$881
& F438W &  1140 \\
& & F555W &  1200 \\
& & F625W &  1134 \\
& & F814W &  1152 \\
\hline
\end{tabular}
\label{obs.tab}
\end{table}

NGC~4666 was observed by the HST as part of the GO program ``Going gently into the night: constraining Type Ia supernova nucleosynthesis using late-time photometry" (ID: 14611; PI: Graur O.). They were conducted by the Wide Field Camera 3 (WFC3) Ultraviolet-Visible (UVIS) channel in four filters (F438W, F555W, F625W, F814W) and at five epochs from Apr~21 to Aug~07 in 2018 (Table~\ref{obs.tab}). We retrieved the images, which have been flattened and corrected for charge transfer efficiency (i.e. \texttt{*\_flc.fits}), from the Mikulski Archive for Space Telescopes\footnote{\url{https://archive.stsci.edu/index.html}} and aligned them with \textsc{tweakreg} based on hundreds of common stars. Images in each band were drizzled together with \textsc{astrodrizzle} using the same parameters as the standard pipeline except \texttt{driz\_cr\_grow~= 3} for better cosmic-ray removal (in Section~\ref{photometry.sec} we shall repeat this step but with finer output scales for better sampling of the sources). We finally performed point-spread-function (PSF) photometry with \textsc{dolphot} \citep{dolphot.ref} on all the images with built-in model PSFs calculated with the \textsc{tinytim} package.

NGC~4666 was also observed by the IFU instrument MUSE \citep{muse.ref} mounted on the Very Large Telescope (VLT) as part of ``The All-weather MUse Supernova Integral field Nearby Galaxies (AMUSING) survey II: ASAS-SN supernova rates with respect to environment properties" (Program ID: 096.D-0296; PI: Anderson~J.). Four individual observations were conducted on Feb 29, Mar 26, and Mar 30 2016, each consisting of four on-source 702-s exposures. The wide-field mode was used and each observation maps a spatial extent of 1~$\times$~1~arcmin$^2$ with 0.2~arcsec sampling. The fields of the four observations are centred on three positions and the combined dataset forms a larger and contiguous mosaic of NGC~4666. The observations cover a wavelength range of 4750--9350~\AA\ with 1.25-\AA\ sampling, and the spectral resolution (R~= d$\lambda$/$\lambda$) ranges from 1770 at the blue end to 3590 at the red end of the spectrum. The raw data have been well processed by the standard pipeline; in brief, all exposures were firstly corrected for instrumental signatures (e.g. bias, flat fields, wavelength calibration, etc.); flux calibration were then performed with standard stars (pre-defined every night) and the sky exposures were used for sky subtraction from the science exposures; finally, the four individual observations were combined to form a deep datacube, which was retrieved from the ESO data archive\footnote{\url{http://archive.eso.org}} and used for this work.

Relative astrometry between the MUSE datacube and the HST images is conducted with 18 common stars in the field. The accuracy corresponds to $\sim$1~WFC3/UVIS pixel or $\sim$0.2 MUSE spaxel.

In this work, we also use the reported light curves of SN~2019yvr as presented in \citetalias{K21}.

\section{SN environment}
\label{env.sec}

\subsection{Maps of stars, gas and extinction}
\label{maps.sec}

\begin{figure*}
\centering
\includegraphics[width=1\linewidth, angle=0]{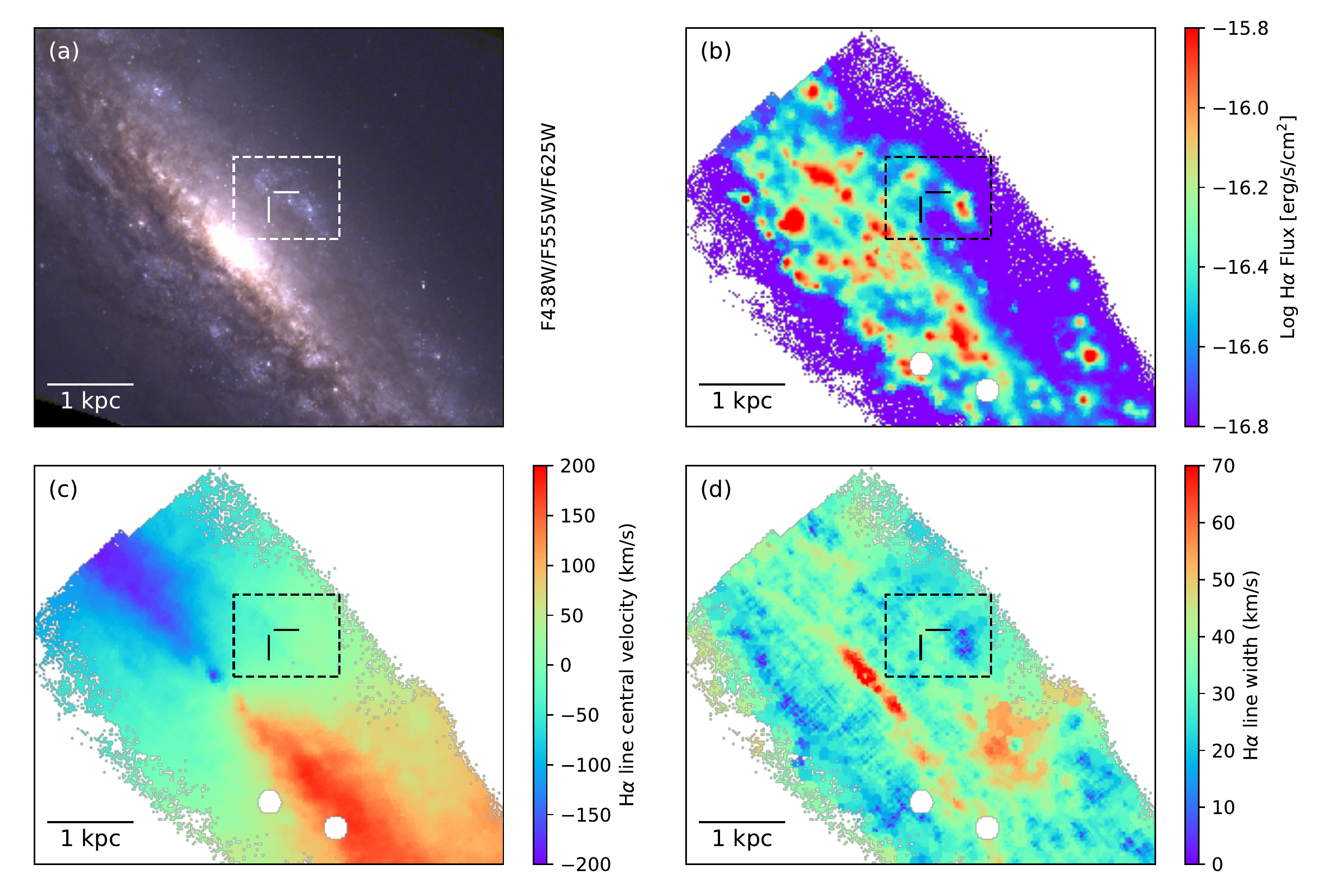} 
\caption{(a) F438W/F555W/F625W three-colour composite image and maps of (b) H$\alpha$ line flux, (c) H$\alpha$ line central velocity and (d) H$\alpha$ line width of SN~2019yvr's host galaxy NGC~4666. In all panels, the crosshair shows the position of SN~2019yvr and the dashed box corresponds to the spatial extent of Fig.~\ref{map2.fig}. All images are oriented with north up and east to the left.}
\label{map1.fig}
\end{figure*}

\begin{figure*}
\centering
\includegraphics[width=0.95\linewidth, angle=0]{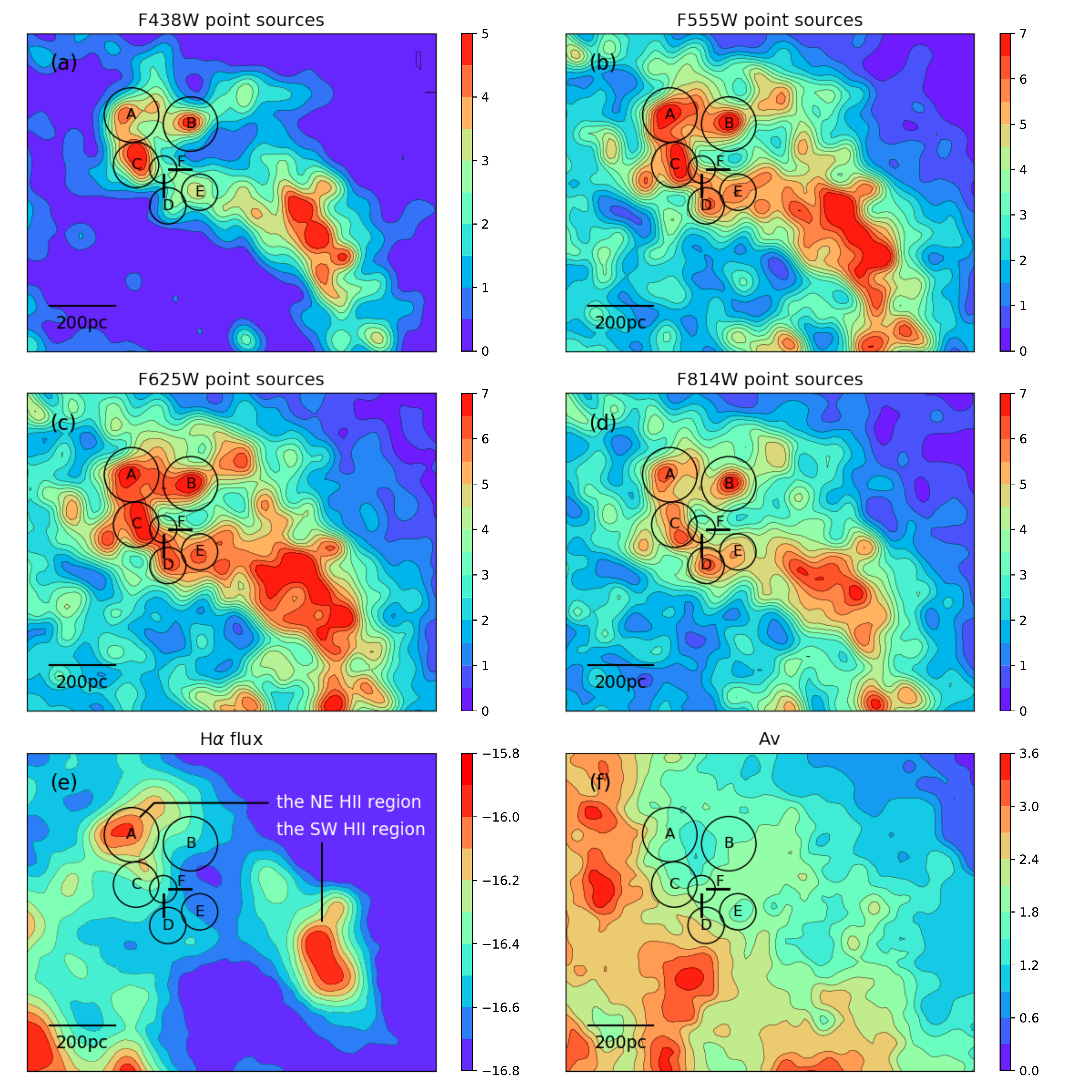}
\caption{Zoom-in views of SN~2019yvr's environment: (a--d) point-source surface densities (in unit of stars~pc$^{-2}$) in the F438W/F555W/F625W/F814W bands; (e) logarithm of H$\alpha$ line flux (in unit of erg~s$^{-1}$~cm$^{-2}$); (f) extinction $A_V$ derived from Balmer decrement (in unit of mag). In all panels, the crosshair shows the position of SN~2019yvr and the circles show \textit{Regions-A} to \textit{-F} for which we derive the stellar ages. All images are oriented with north up and east to the left.}
\label{map2.fig}
\end{figure*}

Figures~\ref{map1.fig} and \ref{map2.fig} show the spatial distributions of stars and gas in the host galaxy and in the local environment of SN~2019yvr, respectively. In the HST three-colour composite image [Fig.~\ref{map1.fig}(a)], a giant star-forming complex is clearly revealed in the vicinity of SN~2019yvr. To analyse the stellar distribution in a quantitative and objective way, we select stars with SNR~$>$~5 in each band from \textsc{dolphot} photometry (Section~\ref{data.sec}) and calculate their surface densities using simple star counts (in spatial bins of 0.2~$\times$~0.2~arcsec$^2$); the derived maps are further smoothed with a Gaussian kernel with a standard deviation of 0.4~arcsec. The surface density maps [Fig.~\ref{map2.fig}(a--d)] show that the star-forming complex is subclustered into many clumps, exhibiting a hierarchical pattern commonly seen in star-forming regions and galaxies \citep[e.g.][]{Sun2017a, Sun2017b, Sun2018}. SN~2019yvr is located away from the surface density peaks and surrounded by five stellar clumps. We define five regions enclosing these clumps with radii of 1.2, 1.2, 1.0, 0.8 and 0.8~arcsec (i.e. 84, 84, 70, 56 and 56~pc), which we shall refer to as \textit{Region-A} to \textit{Region-E}, respectively. We further define a \textit{Region-F} centred on the SN and with a radius of 0.6~arcsec (i.e. 42~pc). We shall use these regions to study the SN environment not only in its immediate surroundings but also on a larger scale in its vicinity.

We used the \textsc{ifuanal} package \citep{ifuanal1.ref, ifuanal2.ref} to analyse the MUSE datacube in order to study the ionised gas in the SN environment with the nebular emission lines. The datacube was de-redshifted, corrected for Galactic reddening, and masked for bright foreground stars. The remaining spaxels were then binned with Voronoi tessellation scheme to achieve a continuum SNR of $\ge$120 in the wavelength range of 5590--5680~\AA. In each bin, the combined spectrum is fitted for its stellar continuum with the \texttt{starlight} package \citep{starlight.ref} based on the \citet{bc03.ref} simple stellar population models. In doing this, we use 15 ages from 3~Myr to 13~Gyr and four metallicities (Z~= 0.004, 0.008, 0.02, 0.05), which correspond to a total of 60 base models. The fitted stellar continuum is scaled to match, and then removed from the observed spectra at the individual spaxels of the Voronoi bin. For the continuum-subtracted datacube, we again create a Voronoi binning of the spaxels so that the combined spectra have a minimum SNR of 5 for H$\beta$. The emission lines in each bin were fitted with Gaussian functions to derive their line flux, central velocity and width (characterised by the standard deviation).

Figure~\ref{map1.fig}(b) shows the map of H$\alpha$ line flux in NGC~4666 and Fig.~\ref{map2.fig}(e) displays a zoom-in view of the SN's local environment. Two giant H~\textsc{ii} regions can be seen in the vicinity of SN~2019yvr, one to its northeast (\textit{NE}) and the other to its southwest (\textit{SW}). Note that the \textit{NE} H~\textsc{ii} region is spatially associated with the stellar clump in \textit{Region-A} and thus probably photo-ionised by the young stars there. SN~2019yvr is located in the outer area of the \textit{NE} H~\textsc{ii} region with a significant offset of $\sim$200~pc from its centre.

The map of H$\alpha$ line central velocity [Fig.~\ref{map1.fig}(c)] shows a clear pattern of galactic disk rotation. The H$\alpha$ line width [Fig.~\ref{map1.fig}(d)] is typically $\sim$30~km~s$^{-1}$ for most regions, except the galaxy's centre and a bubble to the southwest with significantly larger values. Note that an instrumental broadening of $\sim$48.9~km~s$^{-1}$ at the (redshifted) wavelength of H$\alpha$ \citep{lsf.ref} has been removed from the measured line width. There are no special kinematic structures in the local environment of SN~2019yvr.

We further use Balmer decrement to obtain a spatially-resolved extinction map assuming an intrinsic flux ratio of $I$(H$\alpha$)/$I$(H$\beta$)~= 2.87 \citep{agn2.ref} and a \citet{F04.ref} extinction law. Following \citetalias{K21} we adopt a value of the total-to-selective extinction of $R_V$~= 4.7. The derived map is displayed in Fig.~\ref{map2.fig}(f) and the extinction $A_V$ exhibits a significant degree of spatial variation across the field.

\subsection{Metallicity, electron temperature and density}
\label{properties.sec}

\begin{figure*}
\centering
\includegraphics[width=0.8\linewidth, angle=0]{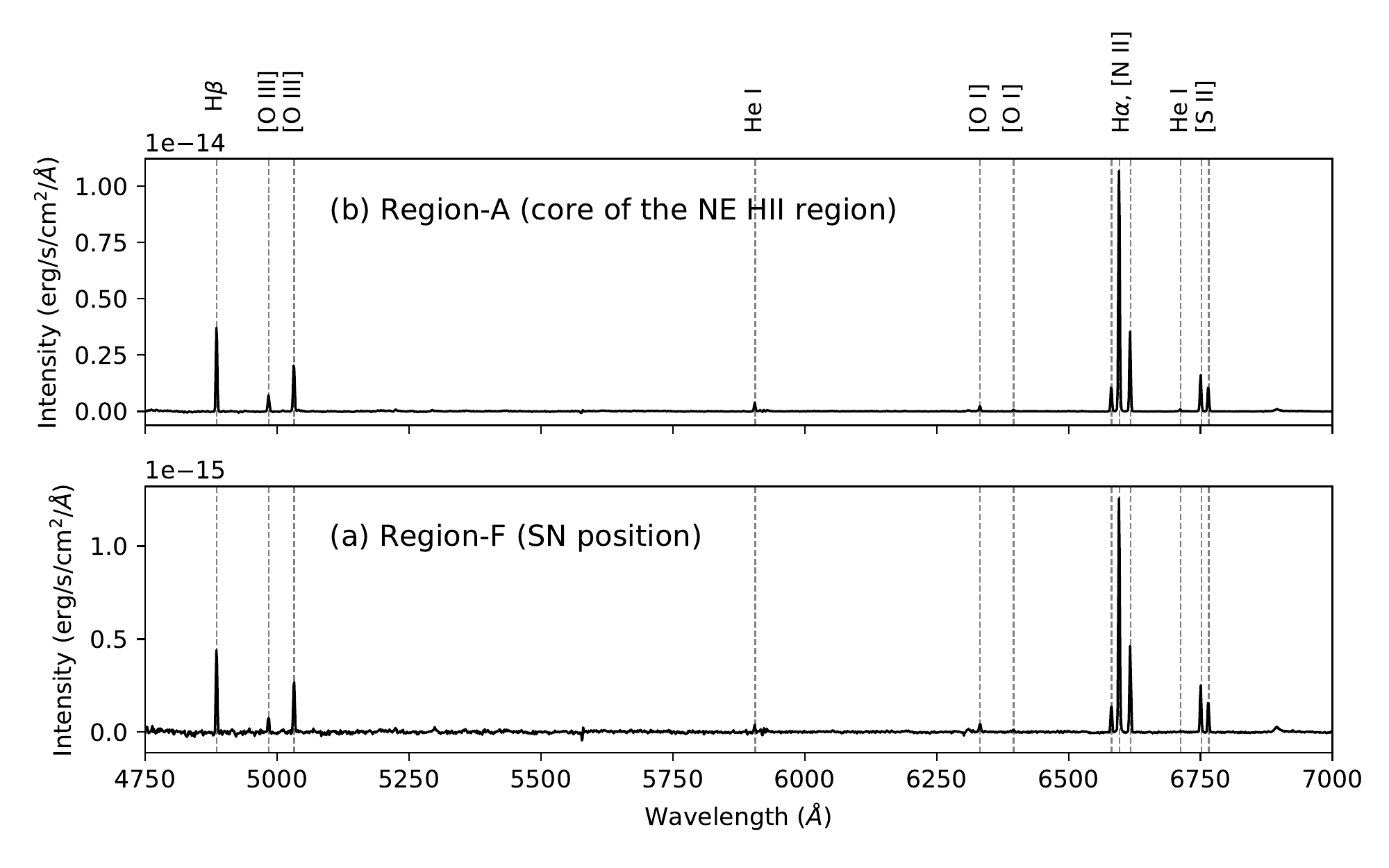}
\caption{Stacked spectra of \textit{Region-A} (top) and \textit{Region-F} (bottom). The stellar continuum has been removed and the interstellar extinction has been corrected.}
\label{spec.fig}
\end{figure*}

We correct each spaxel with the derived interstellar extinction and combine the spaxels within \textit{Region-A} (which encloses the core of the \textit{NE} H~\textsc{ii} region) and \textit{Region-F} (which centres on the SN position) for further analysis. The stacked spectra are shown in Fig.~\ref{spec.fig}, which exhibit a number of important nebular emission lines.

We estimate the gas-phase metallicity with the strong-line method, using the O3N2 calibration based on line ratios of [O~\textsc{iii}]~$\lambda$5007/H$\beta$ and [N~\textsc{ii}]~$\lambda$6584/H$\alpha$ \citep{o3n2.ref}. An oxygen abundance of 12+log(O/H)~= 8.47~$\pm$~0.18~dex is derived, which is slightly lower than the solar value \citep[8.69;][]{solar.ref}; the difference of 0.22~dex is not very significant considering the uncertainties.

We further use the line ratios of [O~\textsc{i}] ($\lambda$6300+$\lambda$6363)/$\lambda$5577 and [N~\textsc{ii}] ($\lambda$6548+$\lambda$6584)/$\lambda$5755 to constrain the electron temperature. Neither [O~\textsc{i}] $\lambda$5577 nor [N~\textsc{ii}] $\lambda$5755 is detected and the detection limits constrain $T_e$~$\lesssim$ 6200~K for \textit{Region-A} and $T_e$~$\lesssim$ 8500~K for \textit{Region-F}. Based on the [S~\textsc{ii}]~$\lambda$6716/$\lambda$6731 diagnostics, we find the electron density to be very low with $n_e$~$\lesssim$~10~cm$^{-3}$ in both regions.

\subsection{Fitting of resolved stellar populations}
\label{pop.sec}

\begin{figure*}
\centering
\includegraphics[width=0.90\linewidth, angle=0]{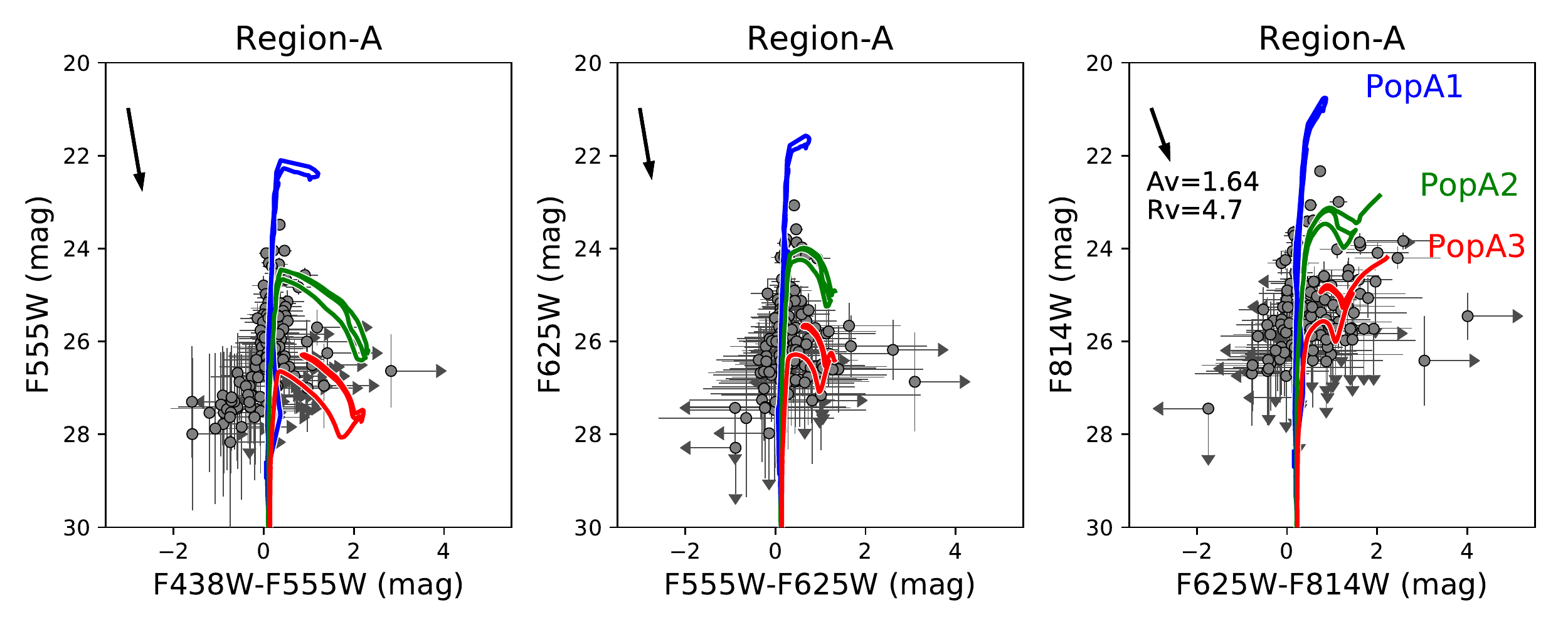}
\includegraphics[width=0.90\linewidth, angle=0]{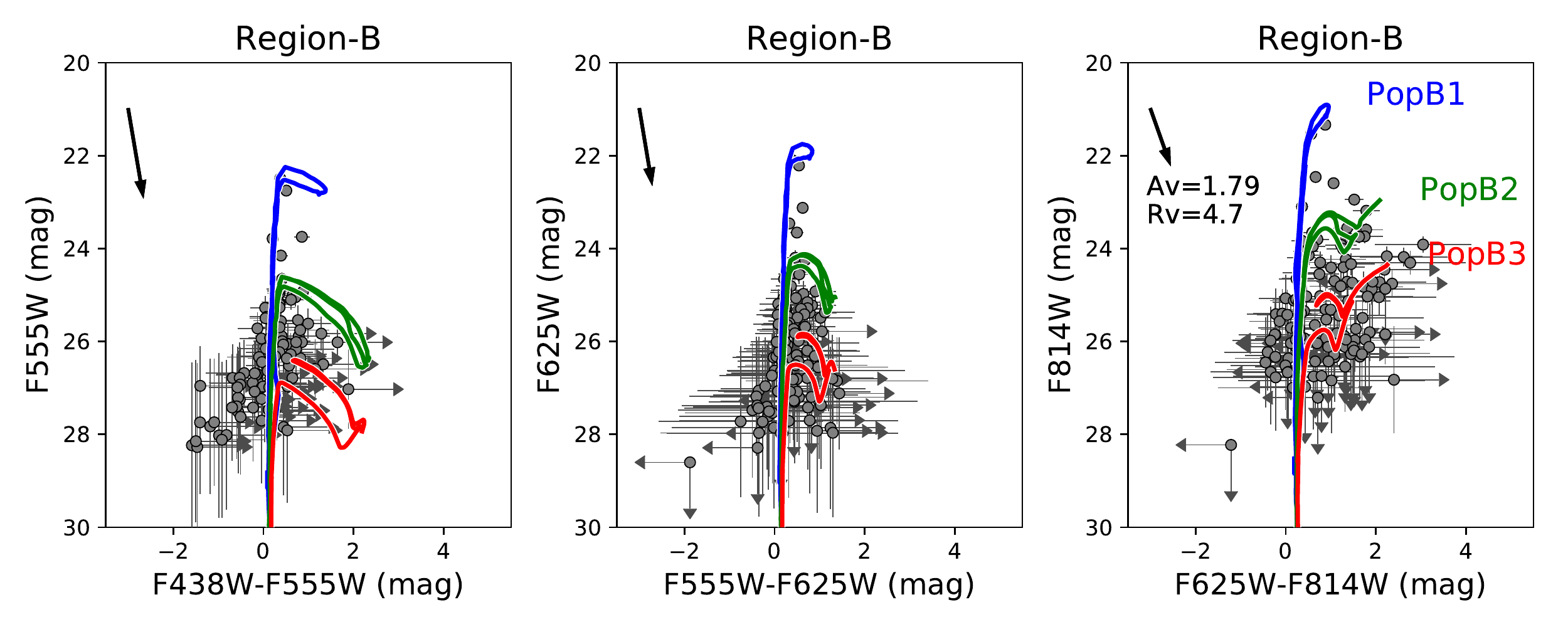}
\includegraphics[width=0.90\linewidth, angle=0]{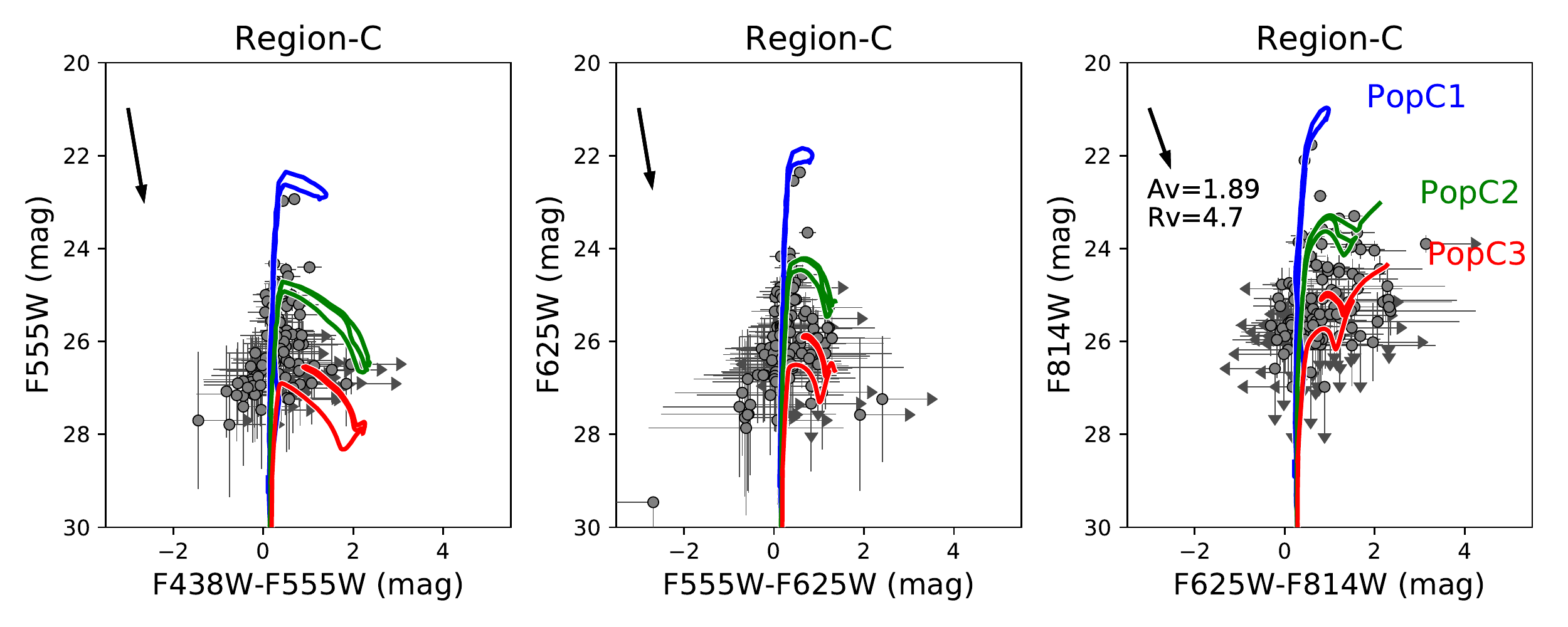}
\caption{CMDs of stars in \textit{Region-A} (top panels), \textit{Region-B} (middle panels) and \textit{Region-C} (bottom panels). In all panels, the blue, green and red isochrones correspond to \textit{PopX1}, \textit{PopX2} and \textit{PopX3} (\textit{X}~= \textit{A}, \textit{B}, \textit{C}), respectively; the arrows are reddening vectors for $R_V$~= 4.7 and for the best-fitting extinctions in each region. See Table~\ref{pop.tab} for a list of the derived ages and extinctions. Note that data points away from the isochrones could be due to the photometric uncertainties, extinction dispersion, and/or binarity.}
\label{cmd1.fig}
\end{figure*}

\begin{figure*}
\centering
\includegraphics[width=0.90\linewidth, angle=0]{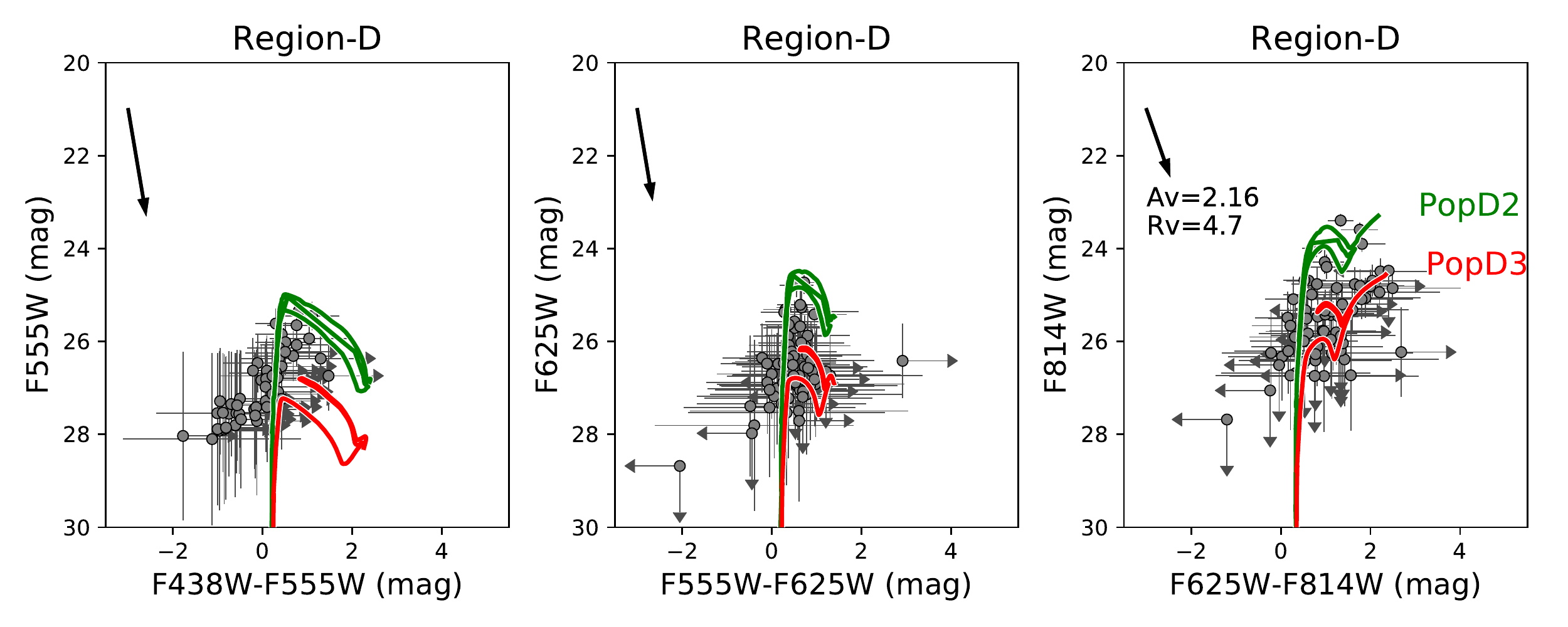}
\includegraphics[width=0.90\linewidth, angle=0]{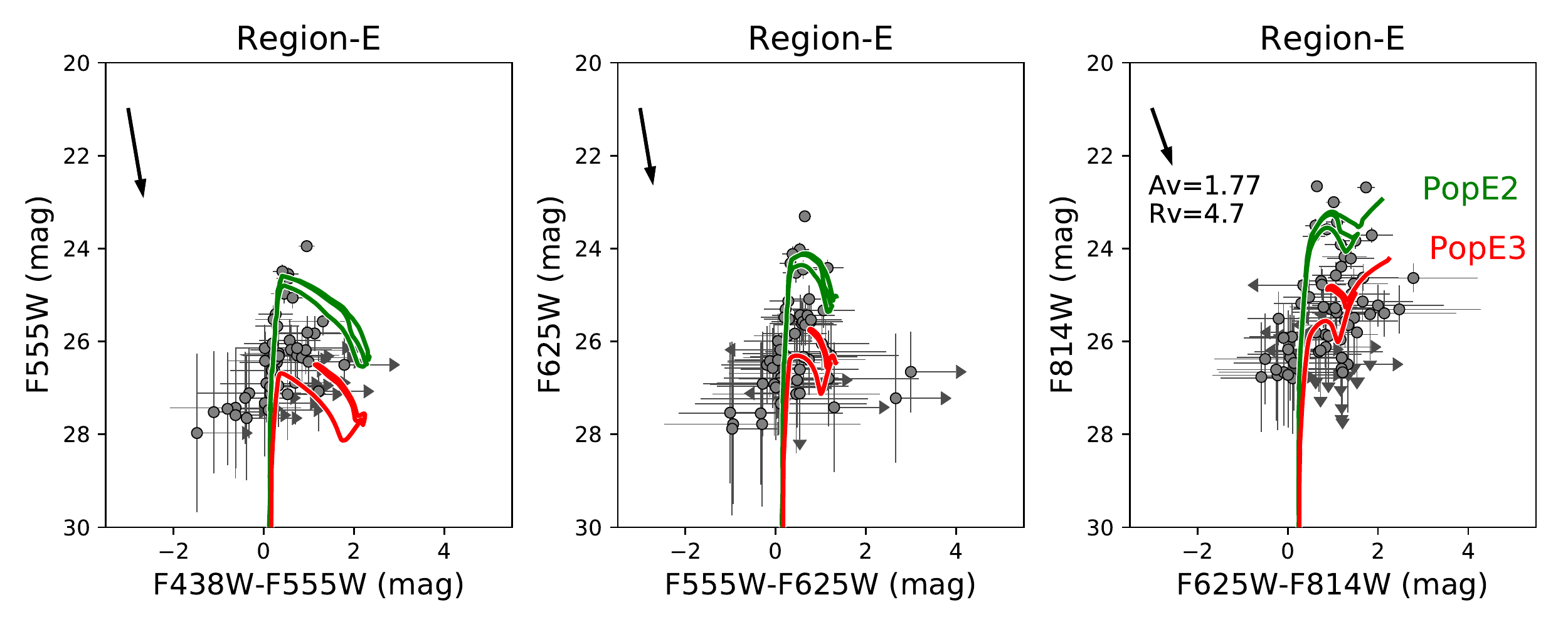}
\includegraphics[width=0.90\linewidth, angle=0]{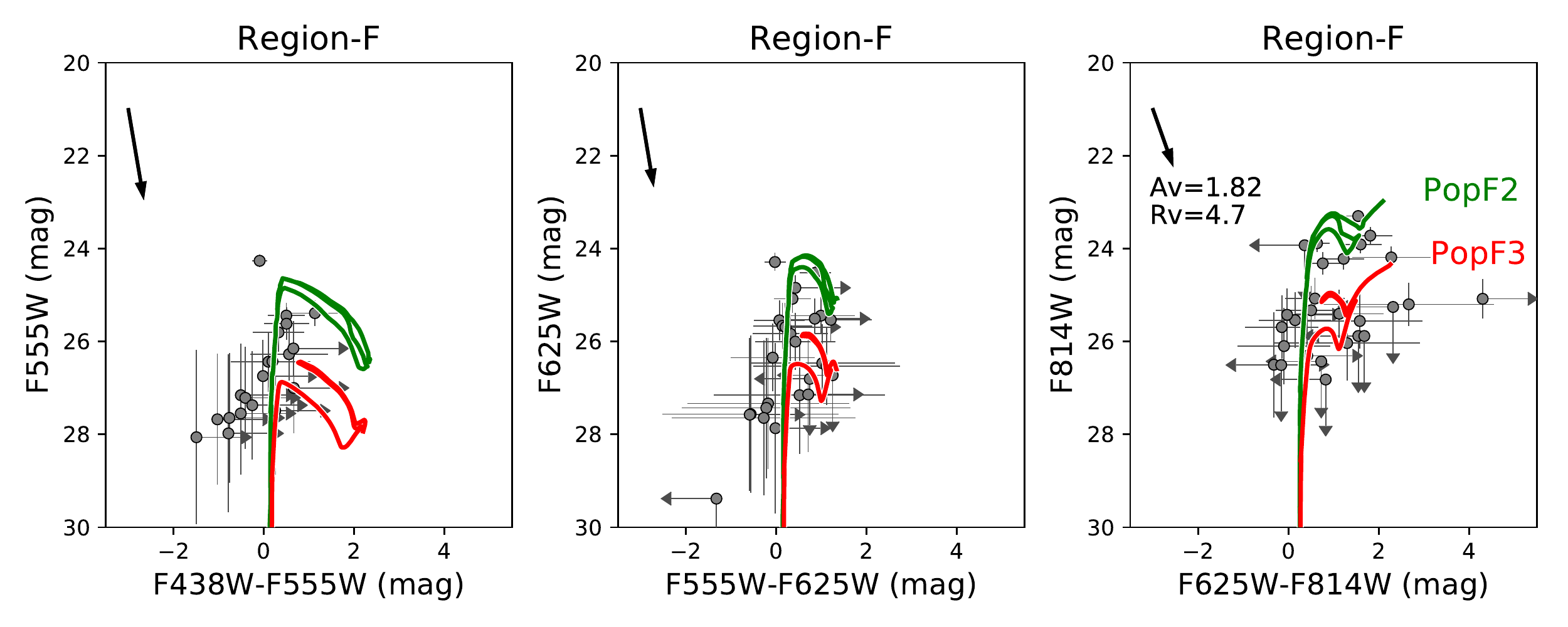}
\caption{Colour-magnitude diagrams of stars in \textit{Region-D} (top panels), \textit{Region-E} (middle panels) and \textit{Region-F} (bottom panels). In all panels, the green and red isochrones correspond to \textit{PopX2} and \textit{PopX3} (\textit{X}~= \textit{D}, \textit{E}, \textit{F}), respectively; the arrows are reddening vectors for $R_V$~= 4.7 and for the best-fitting extinctions in each region. See Table~\ref{pop.tab} for a list of the derived ages and extinctions. Note that data points away from the isochrones could be due to the photometric uncertainties, extinction dispersion, and/or binarity.}
\label{cmd2.fig}
\end{figure*}

\begin{table*}
\caption{Stellar population fitting results. The values in brackets correspond to the parameters' fitting errors. The last column lists the extinction standard deviation of each region, which are derived from the extinction map and fixed in the fitting.}
\begin{tabular}{ccccccc}
\hline
\hline
& Region
& log($t_1$/yr) & log($t_2$/yr) & log($t_3$/yr)
& $A_V$ & $\sigma$$A_V$ \\
& ID
& (\textit{PopX1}) & (\textit{PopX2}) & (\textit{PopX3})
& (mag) & (mag) \\
\hline
$R_V$~= 3.1
& \textit{A}
&  6.61 (+0.02/$-$0.02) &  7.01 (+0.03/$-$0.03) &  7.38 (+0.03/$-$0.03) &  1.16 (+0.04/$-$0.04)
& 0.21 \\
& \textit{B}
&  6.60 (+0.02/$-$0.01) &  7.02 (+0.01/$-$0.01) &  7.41 (+0.02/$-$0.01) &  1.39 (+0.04/$-$0.03) 
& 0.15 \\
& \textit{C}
&  6.62 (+0.02/$-$0.02) &  7.02 (+0.03/$-$0.02) &  7.41 (+0.01/$-$0.02) &  1.42 (+0.04/$-$0.04)
& 0.21 \\
& \textit{D} & --
&  7.05 (+0.03/$-$0.03) &  7.43 (+0.01/$-$0.02) &  1.55 (+0.04/$-$0.05)
& 0.17 \\
& \textit{E} & --
&  7.01 (+0.02/$-$0.02) &  7.40 (+0.02/$-$0.02) &  1.40 (+0.02/$-$0.02)
& 0.09 \\
& \textit{F} & --
&  7.09 (+0.04/$-$0.05) &  7.43 (+0.03/$-$0.04) &  1.21 (+0.02/$-$0.02)
& 0.07 \\
\hline
$R_V$~= 4.7
& \textit{A}
&  6.60 (+0.01/$-$0.01) &  7.00 (+0.03/$-$0.03) &  7.37 (+0.02/$-$0.03) &  1.64 (+0.04/$-$0.04)
& 0.31 \\
& \textit{B}
&  6.61 (+0.02/$-$0.02) &  7.00 (+0.02/$-$0.01) &  7.40 (+0.02/$-$0.02) &  1.79 (+0.04/$-$0.04)
& 0.22 \\
& \textit{C}
&  6.61 (+0.02/$-$0.02) &  7.00 (+0.02/$-$0.02) &  7.38 (+0.02/$-$0.03) &  1.89 (+0.05/$-$0.05)
& 0.30 \\
& \textit{D} & --
&  7.02 (+0.02/$-$0.02) &  7.39 (+0.02/$-$0.03) &  2.16 (+0.05/$-$0.05)
& 0.25 \\
& \textit{E} & --
&  7.00 (+0.02/$-$0.02) &  7.36 (+0.03/$-$0.04) &  1.77 (+0.04/$-$0.04)
& 0.13 \\
& \textit{F} & --
&  7.00 (+0.03/$-$0.03) &  7.39 (+0.04/$-$0.05) &  1.82 (+0.05/$-$0.04) 
& 0.10 \\
\hline
$R_V$~= 6.0
& \textit{A}
&  6.59 (+0.01/$-$0.01) &  6.98 (+0.03/$-$0.03) &  7.36 (+0.03/$-$0.03) &  2.02 (+0.04/$-$0.04)
& 0.39 \\
& \textit{B}
&  6.65 (+0.02/$-$0.03) &  6.99 (+0.02/$-$0.02) &  7.39 (+0.02/$-$0.02) &  2.07 (+0.05/$-$0.04) 
& 0.28 \\
& \textit{C}
&  6.60 (+0.01/$-$0.01) &  6.97 (+0.02/$-$0.02) &  7.34 (+0.03/$-$0.03) &  2.30 (+0.04/$-$0.04)
& 0.38 \\
& \textit{D} & --
&  7.00 (+0.02/$-$0.03) &  7.25 (+0.05/$-$0.03) &  2.69 (+0.04/$-$0.05)
& 0.31 \\
& \textit{E} & --
&  6.99 (+0.02/$-$0.02) &  7.40 (+0.03/$-$0.04) &  2.25 (+0.05/$-$0.05)
& 0.16 \\
& \textit{F} & --
&  6.93 (+0.02/$-$0.03) &  7.28 (+0.07/$-$0.04) &  2.33 (+0.05/$-$0.05)
& 0.12 \\
\hline
\end{tabular}
\label{pop.tab}
\end{table*}

We construct stellar catalogues in \textit{Regions-A} to \textit{-F} based on the \textsc{dolphot} photometry of the HST images in order to quantitatively analyse the stellar populations in the SN environment. In each band, we consider a star to be detected if its recovered parameters meet the following criteria:

(1) type of source, TYPE~= 1;

(2) photometry quality flag, FLAG~$\leq$~3;

(3) source crowding, CROWD~$\leq$~2;

(4) source sharpness, $-$0.5~$\leq$~SHARP~$\leq$~0.5;

(5) signal-to-noise ratio, SNR~$\geq$~5.

\noindent
For stars that are not detected in a filter, we use artificial star tests to determine the detection limit at their positions; an artificial star is considered to be successfully recovered if it is found within 1~pixel of the inserted position and its \textsc{dolphot} parameters meet all the above criteria. We also use artificial stars, randomly positioned in \textit{Regions-A} to \textit{-F}, to determine additional photometric uncertainties due to, e.g., source crowding and large sky background gradient. Note that we exclude SN~2019yvr's pre-explosion source from the stellar catalogues.

Figures~\ref{cmd1.fig} and \ref{cmd2.fig} show the colour-magnitude diagrams (CMDs) of the observed stars in the six regions. We try to fit the stars with model stellar populations (based on \textsc{parsec} stellar isochrones; \citealt{parsec.ref}) using a hierarchical Bayesian approach; each model population has a \citet{imf.ref} initial mass function, a 50\% binary fraction, and a flat distribution of primary-to-secondary mass ratio (for simplicity, all binaries are assumed to be non-interacting binaries; see \citealt{Maund2016a} and \citealt{Sun2021} for a detailed description of the method). It has been shown that a model population can be considered as a burst of star formation with a small age spread \citep[prolonged star formation can be considered as the mixture of many bursts;][]{Walmswell2013, Maund2016a}. Thus, we assume each model population to have a Gaussian distribution of stellar log-age with a small standard deviation of 0.05~dex. The stellar extinctions are also assumed to follow a Gaussian distribution, for which priors on the mean value and the standard deviation are found using the extinction map derived with Balmer decrement (Section~\ref{maps.sec}); the standard deviation is fixed in the fitting for simplicity while the mean value is allowed to vary around the prior value with a typical uncertainty of 0.05~mag. We first use a total-to-selective extinction of $R_V$~= 4.7 (the best-fitting value found by \citetalias{K21}) and solve for the posterior probability distributions; we then change $R_V$ to 3.1 and 6.0 and repeat the analysis (note that the extinction map from Balmer decrement is also updated) to assess the effect of alternative $R_V$ values.

The derived parameters are listed in Table~\ref{pop.tab} and stellar isochrones that correspond to the best-fitting model populations are overlaid in the CMDs (Figs.~\ref{cmd1.fig}, \ref{cmd2.fig}). We find that stars in \textit{Region-A/B/C} can be fitted with three stellar populations with increasing mean log-ages around 6.6, 7.0, 7.4 [we shall refer to them as \textit{PopX1}, \textit{PopX2} and \textit{PopX3} (\textit{X}~= \textit{A}, \textit{B}, \textit{C}), respectively]. In contrast, \textit{Region-D/E/F} are well fitted with only two stellar populations with mean log-ages of $\sim$7.0 and $\sim$7.4 [\textit{PopX2} and \textit{PopX3} (\textit{X}~= \textit{D}, \textit{E}, \textit{F}) hereafter]. It is worth noting that the ages of \textit{PopX1/2/3} between different regions are very similar to each other; in addition, altering the value of $R_V$ has only a very small effect on the derived ages.

From these results, it is clear that there have been three episodes of star formation occurring at $\sim$4~Myr, $\sim$10~Myr and $\sim$25~Myr ago. Stars that form from the earlier two episodes are distributed across the field while the youngest stars from the most recent episode are concentrated only in \textit{Region-A/B/C}. This is consistent with \textit{Region-A/B/C} being spatially coincident with or very close to the \textit{NE} giant H~\textsc{ii} region; in contrast, \textit{Region-D/E/F} are associated with significantly fainter H$\alpha$ emission, suggesting an older evolutionary stage. This is also supported by the stellar map in the bluest F438W filter, where the surface density peaks are much less pronounced in \textit{Region-D/E/F} compared with those in \textit{Region-A/B/C}.

In the above analysis, interacting binaries were not considered for the model stellar populations. In practice, however, a large fraction of all massive stars will exchange mass with a companion during their lives \citep[e.g.][]{Sana2012}. Mass accretion and mergers in close binaries can create some rejuvenated stars that look younger than their true ages \citep[e.g.][]{Schneider2015}. Therefore, our results tend to underestimate the ages of each stellar population. It is not trivial to model stellar populations containing interacting binaries so we ignore this effect in this work.

In the above analysis we have assumed, in each region, all stars and the ionised gas to have the same extinction distributions. We cannot exclude the possibility that stars of different ages and the ionised gas may lie at different positions along the line of sight and have different (mean) extinctions. Such a phenomenon was found by \citet{Sun2021} in the environment of SN~2004dg and SN~2012P, where the ionised gas, younger stars and older stars have decreasing extinctions and are probably distributed from further to nearer toward the observer, respectively. They suggested this phenomenon is likely due to the spiral arm, which swept through their positions and triggered star formation sequentially at different times. In the case of SN~2019yvr, however, there is no obvious signature of spiral arms in its vicinity [Fig.~\ref{map1.fig}(a)]; instead, it is more likely that, inside the active star-forming complex, the stars and gas are still spatially intermingled and thus have very similar extinction distributions. We have tested the population fitting by removing the prior on the mean stellar extinction and leaving it as a free parameter; the fitted values are still very close to those from Balmer decrement.

It is also interesting to compare the extinction of \textit{Region~F} with that of SN~2019yvr itself. \citetalias{K21} derived $A_V$~= 2.4$^{+0.7}_{-1.1}$ and $R_V$~=~4.7$^{+1.3}_{-3.0}$ by fitting templates to the SN's colour curves [see their Fig.~5 for the probability distribution on the ($R_V$, $A_V$) plane]. Our derived extinctions for \textit{Region~F} are close to their lower limit but still within uncertainties. It is worth noting that SN~2019yvr may have a higher extinction than the surrounding stars due to circumstellar dust. However, the large extinction uncertainties make it difficult to assess this effect via a simple comparison of the derived extinctions. Also note that the pre-explosion source of SN~2019yvr has been excluded from the CMD fitting so its possible circumstellar extinction does not affect our results.

This analysis suggests that the very local environment of SN~2019yvr, \textit{Region-F}, is dominated by two stellar populations, \textit{PopF2} with a mean log-age of log($t$/yr)~$\sim$~7.0~$\pm$~0.1 and \textit{PopF3} with log($t$/yr)~$\sim$~7.4~$\pm$~0.1 (the quoted errors include the effect of extinction law uncertainties). Assuming the progenitor's lifetime is not significantly affected by binary evolution, if any, the SN progenitor would have an initial mass of $M_{\rm ini}$~= 19.7$^{+4.8}_{-3.4}$~$M_\odot$ if it arose from \textit{PopF2} or $M_{\rm ini}$~= 10.4$^{+1.5}_{-1.3}$~$M_\odot$ if it had the same age as \textit{PopF3} (according to the \textsc{parsec} stellar models; \citealt{parsec.ref}).

\section{Bolometric light curve}
\label{curve.sec}

\begin{figure}
\centering
\includegraphics[width=1.0\linewidth, angle=0]{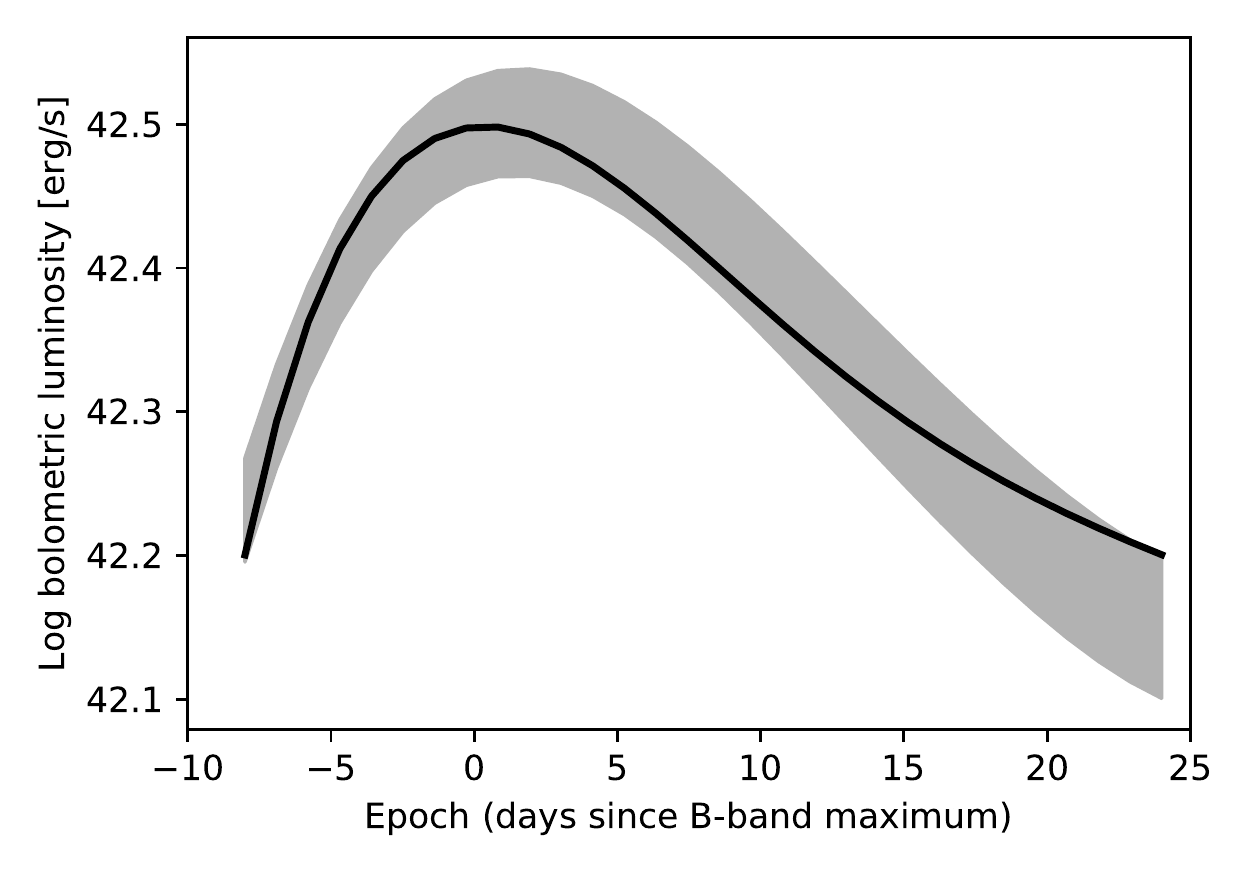}
\caption{The bolometric light curve of SN~2019yvr derived from the reported $gri$ photometry of \citetalias{K21}. The grey-shaded region reflects its uncertainties propagated from errors in the photometry and the bolometric correction. The solid line is the best-fitting \citet{Arnett1982} model.}
\label{curve.fig}
\end{figure}

To help distinguish which stellar population (\textit{PopF2} or \textit{PopF3}) is more likely to have the same age as SN~2019yvr's progenitor, we try to derive the SN ejecta mass by modelling its bolometric light curve. We use the observational data from \citetalias{K21} (their Table~A.1), converting the $g$-, $r$-, $i$-band magnitudes from the Pan-STARRS photometric system to SDSS system \citep{Finkbeiner2016}, and correcting them for interstellar extinction using values of $A_V$~= 2.4 and $R_V$~= 4.7 \citepalias{K21}. We then fit low-order polynomials to the light curves and resample them with a regular spacing in epoch. Bolometric corrections are calculated from the $g - r$ and $g - i$ colours based on the calibration of \citet{Lyman2014} and applied to the $g$-band magnitudes. The results from the $g - r$ and $r - i$ colours are very similar to each other and we take their average as the final bolometric light curve (Fig.~\ref{curve.fig}). An \citet{Arnett1982} model is fitted to the curve (using the same method and parameters as in \citealt{Lyman2016}), which yields a $^{56}$Ni mass of 0.18~$M_\odot$ and the parameter $\tau_m$~= 13.0~days.

The ejecta mass $M_{\rm ej}$ is related to the parameter $\tau_m$ by the equation
\begin{equation}
\tau_m^2 = \dfrac{2 \kappa_{\rm opt} M_{\rm ej} }{\beta c v_{\rm sc}},
\end{equation}
where $\kappa_{\rm opt}$ is the opacity (0.06~cm$^2$~g$^{-1}$; \citealt{Maeda2003}), $\beta$~$\sim$~13.8 a constant, $c$ the speed of light, and $v_{\rm sc}$ the SN's scale velocity that is observationally set as the photospheric velocity at maximum light. We estimate $v_{\rm sc}$ with the Fe~\textsc{ii}~$\lambda$4924, Fe~\textsc{ii}~$\lambda$5018 and Fe~\textsc{ii}~$\lambda$5169 features in two spectra of SN~2019yvr taken on 29~Dec 2019 (retrieved from the Transient Name Server\footnote{\url{https://www.wis-tns.org/}}). A velocity of $\sim$10~000~km~s$^{-1}$ is obtained. Note, however, this value overestimates the scale velocity since the spectra were observed at $\sim$7 days before $V$-band maximum; yet this value is consistent with the mean photospheric velocity for Type~Ib SNe \citep{Branch2002} at similar epochs (within typical errors of 1000~km~s$^{-1}$ in the velocity and a few days in the dates of maximum light). \citet{Branch2002} found the photospheric velocity to follow a power-law decline and the mean value drops to $\sim$9000~km~s$^{-1}$ near peak. If we adopt this velocity as SN~2019yvr's scale velocity, an ejecta mass of $M_{\rm ej}$~= 2.0~$M_\odot$ can be derived, consistent with the typical values for Type~Ib SNe \citep{Drout2011, Taddia2015, Taddia2018, Lyman2016, Prentice2016, Prentice2019}.

The estimate of $M_{\rm ej}$ is subject to uncertainties from a number of sources. The uncertainty propagated from errors in the photometry and the bolometric correction corresponds to $\sim$0.1~$M_\odot$ in the derived ejecta mass. A typical error of 1000~km~s$^{-1}$ in the scale velocity would further induce an uncertainty of $\sim$0.2~$M_\odot$ in $M_{\rm ej}$. We also repeated the above analysis with different values of $A_V$ and $R_V$ drawn from the probable region on the $A_V$--$R_V$ plane as found by \citetalias{K21}; the values of $M_{\rm ej}$ may change by $\sim$0.3~$M_\odot$ due to the uncertainty in the extinction and extinction law.

Despite all these uncertainties, the derived ejecta mass is low and consistent with a moderately massive progenitor. Before its final explosion, the progenitor should be a bare helium core of not more than 4.0~$M_\odot$ if we assume a typical mass of 1.4~$M_\odot$ for the compact remnant \citep{ns.ref} and take 2.6~$M_\odot$ as a conservative upper limit for the ejecta mass (taking all the above-mentioned uncertainties into consideration). We checked the Type~Ib SN progenitors in the \textsc{bpass} \citep{bpass.ref} and \citet{Yoon2017} (hereafter \citetalias{Yoon2017}) models (with solar metallicity); no progenitors with initial masses of $\ge$16~$M_\odot$ can evolve to such a low-mass helium star and produce a SN with such a low ejecta mass. \citetalias{K21} mentioned that SN~2019yvr was surrounded by some dense CSM, suggesting that the progenitor has shed a significant amount of material at $\sim$40 years before its explosion; this process is not considered in the \citetalias{Yoon2017} or \textsc{bpass} models. Unlike the hydrogen-poor ejecta, however, the CSM was hydrogen-rich so it originated from the hydrogen envelope, not the helium core, of the progenitor star. Therefore, the derived ejecta mass is still a good indicator of the helium core mass that is directly correlated with the stellar initial mass.

In the previous section we have derived two possible initial masses for SN~2019yvr's progenitor and the above argument disfavours the higher-mass solution, unless a significant amount of material fell back onto the central compact remnant. Thus, the SN progenitor is most likely to have come from the older population \textit{PopF3} and have a lifetime of log($t$/yr)~= 7.4~$\pm$~0.1 and an intial mass of $M_{\rm ini}$~= 10.4$^{+1.5}_{-1.3}$~$M_\odot$. A star of this mass cannot lose its hydrogen envelope via its stellar wind alone \citep{Crowther2007} and was most likely stripped via interaction with a binary companion \citep{Pods1992}.

\section{The Progenitor Detection}
\label{detection.sec}

\citetalias{K21} reports the detection of SN~2019yvr's progenitor (system) candidate in the pre-explosion HST images. Its SED is consistent with a single YHG with effective temperature $T_{\rm eff}$~= 6800~$\pm$~400~K and luminosity log($L$/$L_\odot$)~= 5.3~$\pm$~0.2. This is much cooler and more luminous than a hot and compact helium star as expected for a Type~Ib SN progenitor. In this section we try to explore the nature of this pre-explosion source and the origin of its peculiar SED.

\subsection{Photometry}
\label{photometry.sec}

\begin{figure*}
\centering
\includegraphics[width=1\linewidth, angle=0]{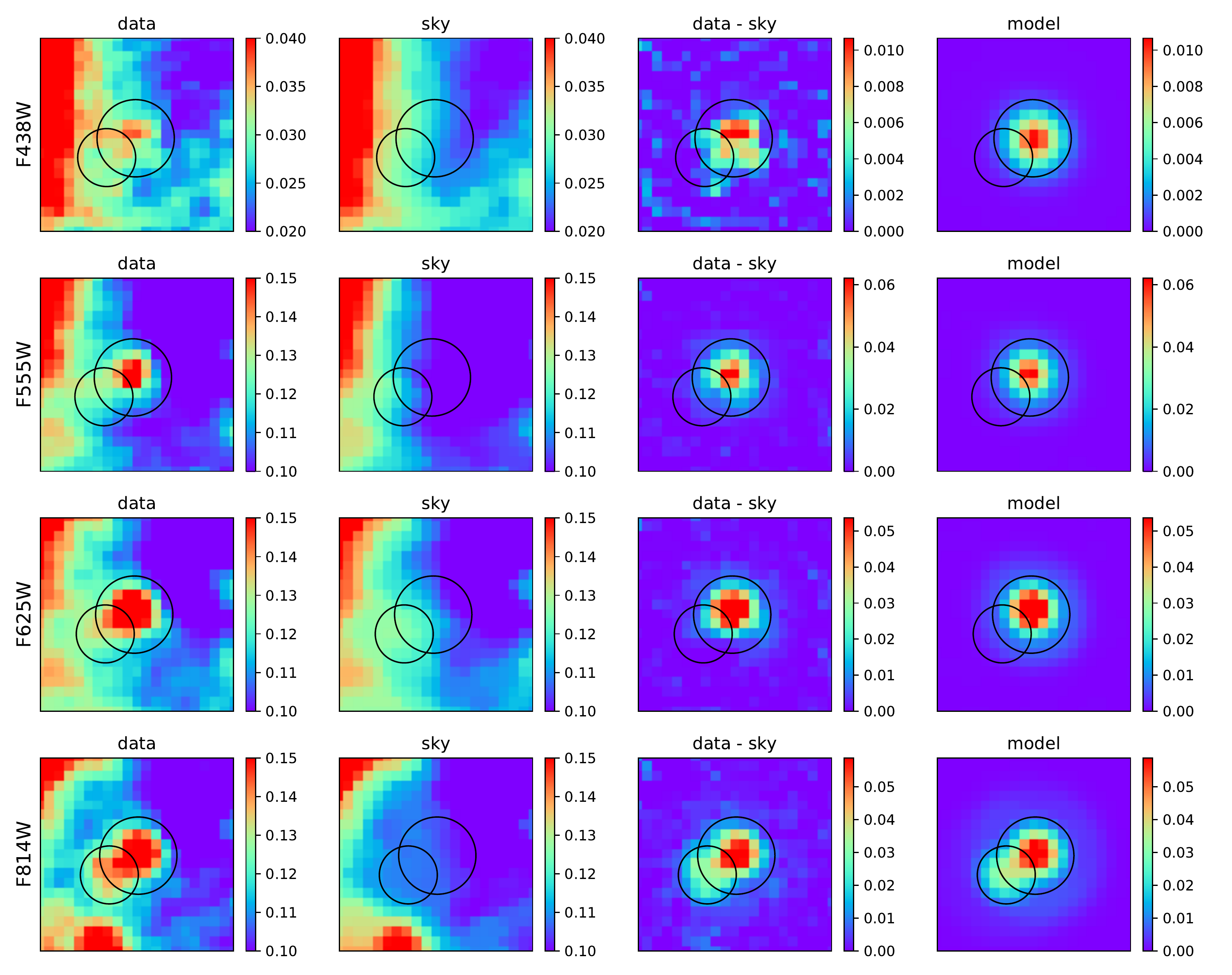}
\caption{Combined images of SN~2019yvr's pre-explosion source (1st column), interpolated sky background (2nd column), sky-cleaned images (3rd column) and the best-fitting source model images (last column). Images of the last three columns are from the last iteration of the fitting (see text). In all panels, the two circles show the mask regions for the progenitor and its close neighbour. The color bars are in units of counts.}
\label{galfit.fig}
\end{figure*}

\begin{table}
\centering
\caption{Magnitudes (in the Vega system) of SN~2019yvr's pre-explosion source obtained by modelling the HST images with \textsc{galfit}.}
\begin{tabular}{ccc}
\hline
\hline
Filter & Magnitude & Uncertainty \\
 & (mag) & (mag) \\
\hline
F438W &  26.22 & 0.06 \\
F555W &  25.44 & 0.02 \\
F625W &  24.68 & 0.02 \\
F814W &  24.05 & 0.02 \\
\hline
\end{tabular}
\label{mag.tab}
\end{table}

In the HST images, the SN's pre-explosion source is subject to significant sky background with a high degree of spatial variation. When we tried to perform its photometry with \textsc{dolphot}, the derived magnitudes are very sensitive to how the sky background is fitted (e.g. fit the sky inside the PSF region but outside the photometry aperture with \texttt{FitSky~= 2} or within the photometry aperture as a 2-parameter PSF fit with \texttt{FitSky~= 3}). The recommended values of the photometry aperture size (\texttt{img\_RAper~= 3} for \texttt{FitSky~= 2} and \texttt{img\_RAper~= 8} for \texttt{FitSky~= 3}; see the \textsc{dolphot} user manual) also seem too large to account for the spatial variation of the sky background. The results with different photometry parameters can differ by a few tenths of magnitude, much larger than the uncertainties reported by \textsc{dolphot}. For statistical studies of stellar populations (e.g. Section~\ref{pop.sec}), this additional error can be estimated by artificial star tests and taken into account in their analysis. For a single important source, however, we expect to overcome the challenge of sky background and to achieve a higher precision in its photometry.

In this work, we use a different approach to measure the brightness by directly modelling the pre-explosion source on the HST images. The SN site was observed by HST at 3/5/5/3 epochs in the F438W/F555W/F625W/F814W bands (Table~\ref{obs.tab}) and, at each epoch, the observations were performed with a 3-point dither pattern. Due to its imperfect pointing accuracy, the telescope's pointing might have some small offsets between the different epochs, leading to an effective 15-point dithering in the F555W/F625W bands and 9-point dithering in the F438W/F814W bands. This allows us to, in each band, drizzle all images together with a finer output pixel scale so that the stellar PSF can be better sampled. In practice, we use an output pixel size of 0.02~arcsec or half of the raw pixel size of WFC3/UVIS. The drizzled images are displayed in the first column of Fig.~\ref{galfit.fig}. Note that, in the F814W band, there is a close neighbouring star in the vicinity of the progenitor (system); this star cannot be found by \textsc{dolphot} since it searches for stars on the individual single-exposure images, which have shallower depths and coarser pixel scales than the combined image.

With the better-sampled drizzled images, we first mask the pre-explosion source and its close neighbour (with radii of 4 and 3 pixels, respectively); the sky background in the masked region is estimated with bi-linear interpolation and 3~pixel~$\times$ 3 pixel median-filter smoothing. We then clean the sky background from the raw image and fit the remaining stellar light with the \textsc{galfit} package \citep{galfit1.ref, galfit2.ref} and \textsc{tinytim} model PSFs. In this process, the sky brightness is overestimated (and the stellar brightness is underestimated) since the PSF wings from the stars actually extend beyond the masked regions. We reduce this effect by removing the best-fitting model stars from the raw image and then repeat the above steps for the new image. This process is repeated a few times and the derived magnitudes converge very rapidly within a few iterations. The last three columns of Fig.~\ref{galfit.fig} show the interpolated sky background, sky-cleaned images, and the best-fitting model images for the stellar sources in the final iteration.

Table~\ref{mag.tab} lists the derived magnitudes. The results are very close to those in \citetalias{K21}, except the F814W band in which the magnitude differs by 0.22~mag (note that \citetalias{K21} used the AB magnitude system). This is possibly due to the influence of the close neighbouring star, which has a magnitude of 24.88~$\pm$ 0.04 in the F814W band but not significantly detected in the other bands.

\subsection{SED}
\label{sed.sec}

\begin{figure*}
\centering
\includegraphics[width=0.85\linewidth, angle=0]{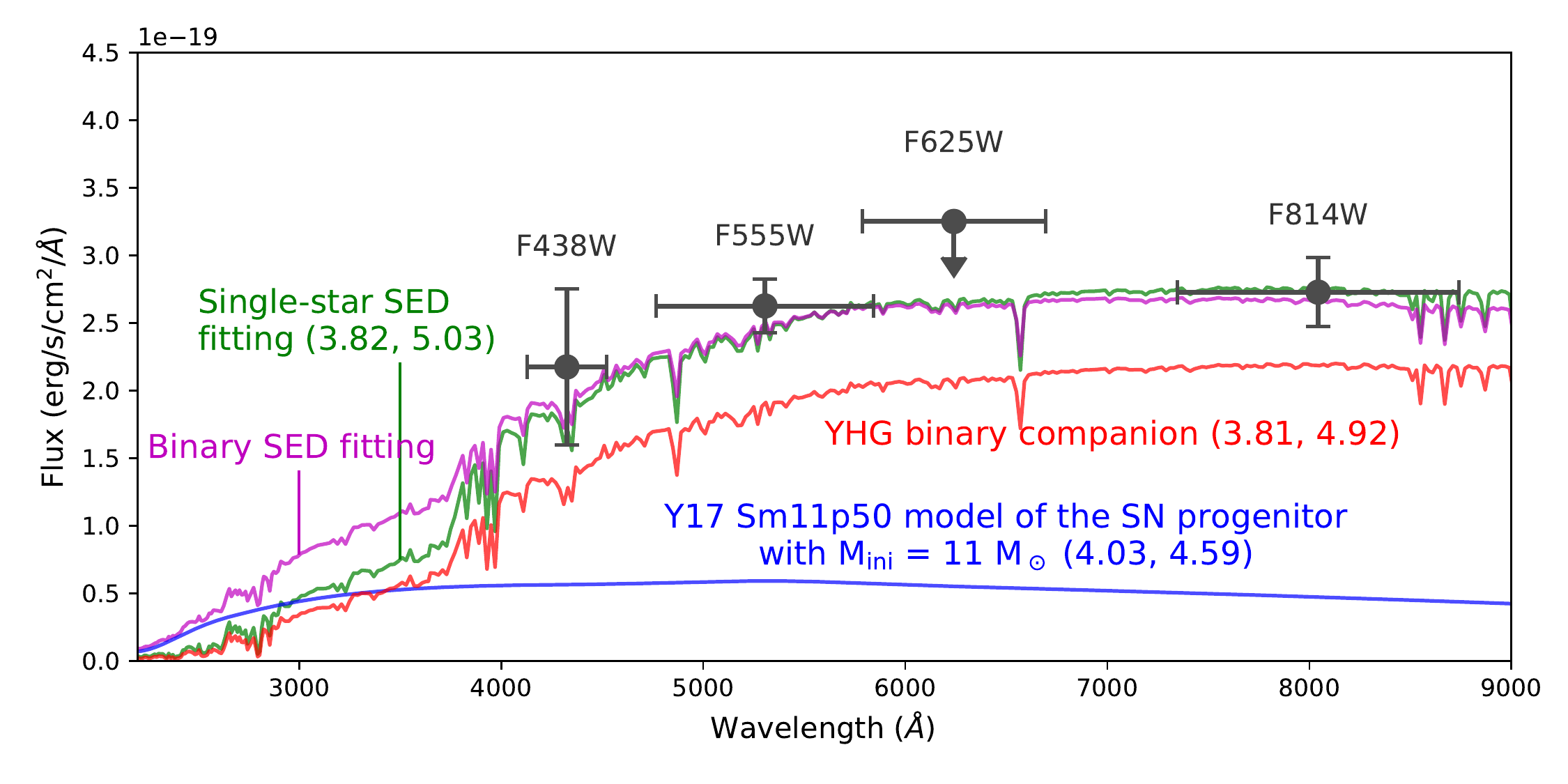}
\caption{SED of SN~2019yvr's pre-explosion source (points); the horizontal error bars are the root-mean-square bandwidths of the filters while the vertical error bars reflect the 5$\sigma$ photometric uncertainties; the F625W magnitude is regarded as an upper limit since it may be contaminated by H$\alpha$ emission. The green/magenta lines correspond to model spectra from single-star/binary SED fitting with $R_V$~= 4.7 and $A_V$~= 2.4, and the blue/red lines are for the progenitor star/binary companion in the binary SED model (i.e. the magenta line is the sum of the blue and red lines). Numbers in the brackets are the stellar effective temperature and luminosity (in logarithm) of the corresponding component.}
\label{sed.fig}
\end{figure*}

\begin{table}
\centering
\caption{Results of single-star SED fitting of SN~2019yvr's pre-explosion source. The 1st column lists the total-to-selective extinctions used in the fitting and the remaining columns are the best-fitting extinction and the star's effective temperature and luminosity. The values in brackets correspond to the parameters' fitting errors.}
\begin{tabular}{ccccccc}
\hline
\hline
$R_V$
& $A_V$/mag & log($T_{\rm eff}$/K) & log($L$/$L_\odot$) \\
\hline
3.1
&  1.7 (+0.1/$-$0.1) &  3.82 (+0.01/$-$0.01) &  4.77 (+0.08/$-$0.08) \\
4.7
&  2.4 (+0.2/$-$0.2) &  3.82 (+0.02/$-$0.02) &  5.03 (+0.10/$-$0.10)  \\
6.0
&  2.6 (+0.3/$-$0.2) &  3.81 (+0.02/$-$0.01) &  5.12 (+0.11/$-$0.11) \\
\hline
\end{tabular}
\label{sedsin.tab}
\caption{Results of binary SED fitting of SN~2019yvr's pre-explosion source. The 1st column lists the total-to-selective extinctions used in the fitting and the remaining columns are the best-fitting extinction and the companion star's effective temperature and luminosity. The values in brackets correspond to the parameters' fitting errors. The primary star is simulated with a blackbody with fixed effective temperature log($T_{\rm eff, 1}$/K)~= 4.03 and luminosity log($L_1$/$L_\odot$)~= 4.59 (from the Sm11p50 model of \citetalias{Yoon2017}).}
\begin{tabular}{ccccccc}
\hline
\hline
$R_V$
& $A_V$/mag & log($T_{\rm eff, 2}$/K) & log($L_2$/$L_\odot$)\\
\hline
3.1
&  1.7 (+0.1/$-$0.1) &  3.77 (+0.02/$-$0.03) &  4.51 (+0.14/$-$0.15)  \\
4.7
&  2.4 (+0.2/$-$0.2) &  3.80 (+0.02/$-$0.03) &  4.91 (+0.14/$-$0.14)  \\
6.0
&  2.6 (+0.3/$-$0.3) &  3.80 (+0.02/$-$0.03) &  5.02 (+0.14/$-$0.14)  \\
\hline
\end{tabular}
\label{sedbin.tab}
\end{table}

\subsubsection{Single-star SED fitting}
\label{sedsin.sec}

Figure~\ref{sed.fig} displays the SED of SN~2019yvr's pre-explosion source. We firstly fit the observed SED by assuming the brightness is dominated by a single star. In doing this, we use the \textsc{atlas9} model stellar spectra and regard the F625W magnitude as an upper limit since it may be contaminated by H$\alpha$ emission, if any. For $R_V$~=~4.7, we derive an extinction $A_V$~= 2.4$^{+0.2}_{-0.2}$, a stellar effective temperature log($T_{\rm eff}$/K)~= 3.82$^{+0.02}_{-0.02}$, and luminosity log($L$/$L_\odot$)~= 5.03$^{+0.10}_{-0.10}$ (results for three different $R_V$ values are listed in Table~\ref{sedsin.tab}).

The above results agree with \citetalias{K21} within uncertainties; and as they pointed out, the low effective temperature and high luminosity are not expected. In Fig~\ref{sed.fig} we show the model blackbody spectrum (blue line) for a Type~Ib SN progenitor with effective temperature log($T_{\rm eff}$/K)~= 4.03 and luminosity log($L$/$L_\odot$)~= 4.59; these values are from the Sm11p50 model of \citetalias{Yoon2017} and are typical for Type~Ib SN progenitors with an initial mass of 11~$M_\odot$ matching our constraint (10.4$^{+1.5}_{-1.3}$~$M_\odot$; Sections~\ref{pop.sec} and \ref{curve.sec}). It's clear that an expected Type~Ib SN progenitor is much hotter and fainter than the observed SED. Helium stars can be slightly redder with larger mixing lengths or higher metallicities; we tested with various mixing lengths and metallicities using the \textsc{MESA} package \citep{mesa.ref} for helium stars with masses of 1.75--10~$M_\odot$ (note however the upper limit of 4~$M_\odot$ derived from light curve modeling in Section~\ref{curve.sec}); the results are much less luminous and/or much hotter than the derived parameters and none of the models can match SN~2019yvr's pre-explosion source. \citetalias{K21} proposed that the SN progenitor may have a quasi-photosphere due to dense wind or an inflated hydrogen envelope supported by radiation pressure, leading to a lower effective temperature and a higher luminosity than expected; but they point out this is unlikely due to the lack of flash ionization features and early CSM interaction signatures.

\subsubsection{Binary SED fitting}
\label{sedbin.sec}

Here we consider another possibility that the observed SED of SN~2019yvr's pre-explosion source could be a combination of the progenitor star and a binary companion. Following this assumption, we try to fit the observed SED with two components. The first component is simulated with a blackbody with effective temperature log($T_{\rm eff, 1}$/K)~= 4.03 and luminosity log($L_1$/$L_\odot$)~= 4.59 from the Sm11p50 model of \citetalias{Yoon2017} for a typical Type~Ib SN progenitor (using the other models of Sm11p20, Sm11p200 and Sm11p300 does not significantly affect the following results). The second component, for the binary companion, is modelled with the \textsc{atlas9} synthetic stellar spectra \citep{ck.ref} with its effective temperature log($T_{\rm eff, 2}$/K) and luminosity log($L_2$/$L_\odot$) leaving as free parameters. Assuming $R_V$~= 4.7, we derive best-fitting values of $A_V$~= 2.4$^{+0.2}_{-0.2}$, log($T_{\rm eff, 2}$/K)~= 3.80$^{+0.02}_{-0.03}$ and log($L_2$/$L_\odot$)~= 4.91$^{+0.14}_{-0.14}$ (results for three different $R_V$ values are listed in Table~\ref{sedbin.tab}). The companion's effective temperature and luminosity are consistent with a cool and inflated YHG with a stellar radius of $\sim$230~$R_\odot$.

The model spectra of the binary companion (red line) and of the two stars combined together (magenta line) are displayed in Fig.~\ref{sed.fig}. The observed brightness is dominated by the binary companion instead of the SN progenitor itself; the binary companion is $\sim$1.5 times brighter than the progenitor in the F438W band and $\sim$5 times brighter in the F814W band. Only blueward of 3500~\AA\ does the SN progenitor's brightness exceeds that of its binary companion. It is also worth noting that the single-star and binary SED fittings can reproduce the optical-band magnitudes almost equally well; the Bayes factor of the binary model over the single-star model is only 3.2, so the observed SED cannot distinguish the two scenarios. Their difference is significant only in the ultraviolet wavelengths, which unfortunately were not covered by the pre-explosion observations.

\subsection{The YHG binary companion}
\label{companion.sec}

\begin{figure}
\centering
\includegraphics[width=0.95\linewidth, angle=0]{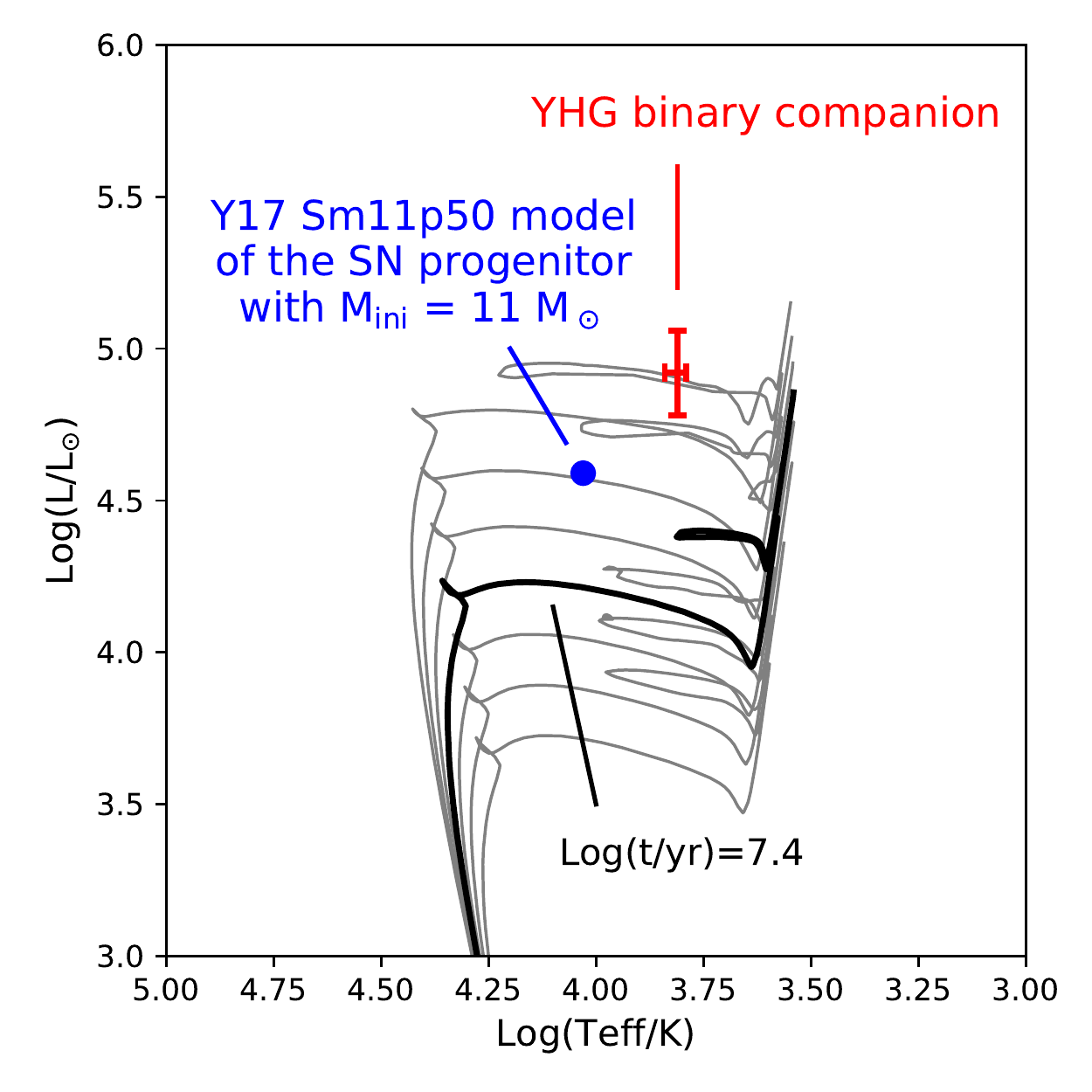}
\caption{Positions of SN~2019yvr's progenitor (blue filled circle) and its binary companion (error bars) on the HR diagram for $R_V$~= 4.7 and $A_V$~= 2.4. The black lines are \textsc{parsec} single-stellar isochrones with ages of log($t$/yr)~= 7.1, 7.2, ..., 7.7 from top to bottom and the thicker one corresponds to the SN progenitor's lifetime as derived from the environmental analysis (Section~\ref{pop.sec}).}
\label{hrd.fig}
\end{figure}

Figure~\ref{hrd.fig} displays the positions of the SN progenitor (the Sm11p50 model of \citetalias{Yoon2017}) and its binary companion on the Hertzsprung-Russell (HR) diagram. It is worth noting that the the companion is located in the Hertzsprung gap, where stars are expected to evolve very rapidly and do not spend much time there. This could be evidence of this star being physically associated with the SN progenitor instead of being an unrelated source in chance alignment (based on the surface density of the nearby sources, \citetalias{K21} estimated a $<$2\% probability for a randomly positioned star to be in chance alignment with SN~2019yvr). The companion also appears far away from the single-stellar isochrone for its age [log($t$/yr)~= 7.4~$\pm$~0.1; Sections~\ref{pop.sec} and \ref{curve.sec}]; thus, its evolution could have been significantly influenced by binary interaction.

Theories predict that, for most binary systems, the companion stars should still reside on the main sequence (MS) at the time of SN explosions of the primary stars (\citealt{Zapartas2017}; see also Fig.~5 of \citealt{Sun2020a} for the prediction of the \textsc{bpass} models). This is because the secondary stars have lower masses and evolve on longer timescales than the primary stars. If any mass transfer takes place when the secondary stars are on the MS, they will be able to adapt their structures, get rejuvenated, and appear to be MS stars of younger ages and higher masses \citep{Pods1992, Langer2012}. It is therefore an intriguing phenomenon that SN~2019yvr's binary companion does not reside on the MS (Fig.~\ref{hrd.fig}).

This phenomenon could arise if the companion had a nearly equal initial mass as the SN progenitor so the two stars have evolved on very similar timescales \citep{Claeys2011, Zapartas2017}. In this case, the companion may have already evolved off the MS and avoided rejuvenation when it accreted material from the progenitor via Roche-lobe overflow; as a result, the companion star became inflated and more luminous due to mass accretion, but it would not return to the MS since it had already established an hydrogen-exhausted core (\citealt{Pods1992}; \citealt{Langer2012}; see also Fig.~2 of \citealt{Claeys2011} for the response of accreting stars at different evolutionary stages). In this scenario, the two stars should have a relatively wide separation since otherwise the reverse mass transfer may lead to a binary merger and/or make the progenitor hydrogen-rich again. The binary population synthesis of \citet{Zapartas2017} found that 0.5--1\% of stripped-envelope SNe may have post-MS companions.

The cool and inflated state of the binary companion could also arise if it is still a MS star but temporarily out of thermal equilibrium at the time of observations. SN~2019yvr showed signatures of CSM interaction at $>$150~days from discovery \citepalias{K21} and the progenitor star may have had a violent eruption shortly ($\sim$40~years) before its SN explosion; if part of the CSM were captured by the companion at high accretion rate, an extended envelope could have been formed around the companion. Alternatively, the SN progenitor and the companion may have experienced very late non-conservative binary interaction, ejecting the CSM. Common-envelope evolution hundreds to several thousand of years before the explosion has been suggested in \citet{Margutti2017} to potentially explain the CSM origin of SN~2014C, without the need of further violent eruption from the SN progenitor. If this kind of very late binary interaction could drive the MS companion out of thermal equilibrium, it could also explain the inflated and cool state of the companion in our study, in case the star can sustain its non-equilibrium state long enough till the pre-explosion observations. However, it is questionable whether a MS companion would be inflated during spiral-in the common envelope and whether the post-common-envelope separation can be kept wide enough to fit a YHG within its Roche lobe.


Compared with the best-fitting model spectrum, the observed SED of SN~2019yvr's pre-explosion source exhibits an excess in the F625W band possibly due to some strong H$\alpha$ emission from the star(s). H$\alpha$ emission may arise from the material ejected during the binary interaction or from the CSM ejected by the progenitor/companion. Note that CSM may exhibit strong near-/mid-infrared emission, but \citetalias{K21} did not detect any such emission in three Spitzer/IRAC bands. Their limits correspond to a CSM mass of $M_{\rm CSM}$~$<$ 9~$\times$ 10$^{-2}$~$M_\odot$ for a dust temperature of $T_{\rm d}$~= 200~K or $M_{\rm CSM}$~$<$ 2~$\times$ 10$^{-6}$~$M_\odot$ for $T_{\rm d}$~= 1500~K, assuming a nominal gas-to-dust ratio of 100. Strictly speaking, we also cannot exclude the possibility of imperfect stellar photometry due to background nebular emission that could be spatially varying on very small scales (Section~\ref{photometry.sec}).

We have performed the SED fitting using different values of $R_V$ (Table~\ref{sedbin.tab}). While the companion's effective temperature remains almost unchanged, its luminosity depends very sensitively on the adopted $R_V$ values. Still, the qualitative conclusion reached above is not affected. Further observations are needed to better constrain the extinction law and the properties of SN~2019yvr's binary companion.

\section{Discussion}
\label{discussion.sec}

\subsection{SNe with companion detections}
\label{discussion_companion.sec}

As mentioned in the Introduction, SN~2019yvr is the second Type~Ib SN (after iPTF13bvn) with a possible detection of its progenitor (system). The analysis presented in this work suggests that its pre-explosion brightness is likely dominated by a cool and inflated YHG companion with some contribution from the hot and compact SN progenitor. This makes SN~2019yvr possibly the fifth core-collapse SN, and perhaps the first Type~Ib/c SN, with a direct detection of its binary companion [after the Type~IIb SN~1993J \citep{Maund2004, Fox2014}, SN~2001ig \citep{Ryder2018} and SN~2011dh \citep{Folatelli2014, Maund2015, Maund2019}, and the Type~Ibn SN~2006jc \citep{Maund2016b, Sun2020a}].

Very interestingly, three out of these five SNe (SN~1993J and SN~2006jc in addition to SN~2019yvr) are possibly from nearly equal-mass binaries in which the companion stars accreted material when they had already evolved off the MS (\citealt{Maund2004, Sun2020a}; Section~\ref{companion.sec}). This could be an observational selection effect since in this case the companion stars are overluminous and more easily observed. We also note that observations have found a statistical excess of nearly equal-mass binaries, known as the so-called ``twin phenomenon" \citep[e.g.][]{ElBadry2019}.

SN~2019yvr provides a precious opportunity to study the progenitor channel of Type~Ib SNe; both the low progenitor mass and the YHG companion suggest significant binary interaction in stripping the progenitor's envelope (Section~\ref{companion.sec}; detailed modelling of the binary evolution is beyond the scope of this work). SN~2019yvr also allows to study the ejecta-companion interaction; since the companion is inflated, a large fraction of its extended envelope may be stripped due to the impact of the SN ejecta \citep{Hirai2014, Hirai2015, Hirai2018, Hirai2020, Sun2020a, Ogata2021}.

\subsection{Metamorphosis from Type~Ib to Type~IIn and the pre-SN mass loss of stripped stars}
\label{loss.sec}

SN~2019yvr showed signatures of narrow H$\alpha$, X-ray, and radio emission at $>$150~days from discovery, suggesting strong interaction between the SN ejecta and dense CSM at these times (\citetalias{K21}). Assuming a SN ejecta velocity of $\sim$10000~km~s$^{-1}$ and a CSM ejection velocity of $\sim$100~km~s$^{-1}$, the CSM is located at a distance of 1.3~$\times$ 10$^{16}$~cm from the progenitor and was ejected at $\sim$40~years before the SN explosion. The H$\alpha$ emission also means that the CSM is hydrogen-rich and chemically different from the hydrogen-poor progenitor star.

This peculiar phenomenon is very similar to that of the famous SN~2014C, which transformed from Type~Ib to Type~IIn (i.e. exhibiting narrow hydrogen lines) on a timescale of one year \citep{Mili2015, Margutti2017}. A handful of stripped-envelope SNe have also been found to exhibit such late-time H$\alpha$ emission [e.g. iPTF13ehe, iPTF15esb, iPTF16bd \citep{Yan2017}; SN2004dk \citep{Mauerhan2018, Pooley2019}; SN~2017dio \citep{K2018}; see also \citealt{Vinko2017}]. This phenomenon may be even more common than those discovered since many SNe have eluded long-term monitoring at very late times.

It is also interesting to note the similarity in the progenitor masses between SN~2019yvr and SN~2014C. By analysing SN~2014C's host star cluster, \citet{Sun2020b} suggested that its progenitor was an $\sim$11-$M_\odot$ star stripped in an interacting binary system. The initial mass is very close to that of SN~2019yvr (10.4$^{+1.5}_{-1.3}$~$M_\odot$; Sections~\ref{pop.sec} and \ref{curve.sec}) and the difference is even smaller than the measurement uncertainties. More studies are still needed to constrain the progenitors for the other SN~2014C-like SNe in order to better understand this special subtype.

For a long time the most massive stars, such as the luminous blue variables (LBVs, with initial masses larger than 25~$M_\odot$; \citealt{Vink2012}), have drawn the most attention in studying the pre-SN mass loss before their final explosions. The progenitor analysis of SN~2014C and the Type~Ibn SN~2006jc/2015G led \citet{Sun2020b} and \citet{Sun2020a} to conclude that much lower-mass stars stripped in binaries can also experience violent outbursts with intensities even matching the LBV giant eruptions. SN~2019yvr provides additional evidence supporting enhanced pre-SN mass loss as an important process not only for the very massive hydrogen-rich stars but also for the hydrogen-poor stars at the low-mass end of core-collapse SN progenitors (see \citealt{Sun2020a} for a detailed discussion).

It is not yet fully clear what physical mechanism(s) drives the SN progenitors' enhanced mass loss just before their core collapse. Current models include, to name a few, convection-excited wave heating \citep{Quataert2012, Shiode2014, Fuller2017, Fuller2018, Wu2021, Leung2021}, continuum-driven super-Eddington winds \citep{Ofek2016}, hydrodynamical instabilities at late nuclear burning stages \citep{Smith2014} and binary merger \citep{Chevalier2012}. \citet{Margutti2017} also proposed that 3.8--10\% of Type~Ib/c SN progenitors can experience very late Case-C common-envelope evolution, which can efficiently eject dense CSM without the need to invoke an extra eruption from the progenitor. Interacting SNe like SN~2019yvr with progenitor constraints will be of vital importance in testing these theoretical models.

\subsection{More possibilities}
\label{possibilities.sec}

\citetalias{K21} found from the \textsc{bpass} models that a $\sim$19~$M_\odot$ + $\sim$1.9~$M_\odot$ close binary system can evolve to an end state matching the HST photometry, in which the progenitor dominates the light and the companion has very little contribution (the scenario presented in this work is not included in the \textsc{bpass} models since it uses a grid spacing of 0.1 in the primary-to-secondary mass ratio and thus misses the nearly equal-mass binaries). As the authors noted, however, the progenitor of this model still retains a residual hydrogen envelope of 0.047~$M_\odot$, conflicting with SN~2019yvr's Type~Ib spectral classification at very early times. Yet, this residual envelope could be ejected at late nuclear burning stages (perhaps by some mechanisms mentioned in the above subsection) after the end of core carbon burning where \textsc{bpass} terminates its calculation \citep{bpass.ref}. In this model, the SN progenitor has a final mass of 7.3~$M_\odot$, which seems too large to be compatible with the low ejecta mass of SN~2019yvr derived from light curve fitting (Section~\ref{curve.sec}); however, we cannot exclude the possibility of significant fallback onto the compact remnant, leading to a low ejecta mass in this case. \citetalias{K21} also discussed the scenarios in which the SN progenitor had a quasi-photosphere due to dense wind or an inflated hydrogen envelope supported by radiation pressure, leading to a lower effective temperature and a higher luminosity (see their Section~5 for a detailed discussion).

In summary, SN~2019yvr serves as an important target to explore the various interesting models of its progenitor (system) and investigate whatever physical processes happened to it before the explosion. Late-time observations, when the SN has faded, will help to confirm the nature of the pre-explosion source. If its brightness were dominated by the progenitor star itself, the source would (almost) completely disappear in the future; if the brightness were dominated by the binary companion (as suggested in this work), the source would only become mildly fainter and we predict its late-time magnitude to be F438W~= 26.7, F555W~= 25.8, F625W~= 25.2 and F814W~= 24.3 (based on the results in Section~\ref{sedbin.sec}). Note, however, there could be significant changes in the companion's brightness if the binary orbit was tight enough such that ejecta-companion interaction becomes important \citep{Hirai2014, Hirai2015, Hirai2018, Hirai2020, Sun2020a, Ogata2021}.


\section{Summary and conclusions}
\label{summary.sec}

In this paper we carry out a detailed analysis of the environment, bolometric light curve, and the progenitor detection for the Type~Ib SN~2019yvr. We reach the following conclusions:

(1) SN~2019yvr occurred in an active star-forming region but shows a significant offset from the peaks of stellar surface density and gaseous nebular emission. There have been three episodes of star formation in the complex at $\sim$4~Myr, $\sim$10~Myr and $\sim$25~Myr ago; the SN's immediate vicinity is dominated by the two older stellar populations.

(2) Fitting of the bolometric light curve obtains a low ejecta mass of only $\sim$2.0~$M_\odot$, suggesting the progenitor most likely from the oldest stellar population derived above. Accordingly, the progenitor has an initial mass of 10.4$^{+1.5}_{-1.3}$~$M_\odot$ and requires binary interaction to strip its hydrogen envelope.

(3) We re-performed photometry of SN~2019yvr's progenitor (system) on the pre-explosion images. Its SED can be reproduced by two components, one for a hot and compact Type~Ib SN progenitor and one for a cool and inflated YHG companion. Assuming $R_V$~= 4.7 and using the Sm11p50 model of \citetalias{Yoon2017} to simulate the SN progenitor, we derive an effective temperature log($T_{\rm eff, 2}$/K)~= 3.80$^{+0.02}_{-0.03}$ and luminosity log($L_2$/$L_\odot$)~= 4.91$^{+0.14}_{-0.14}$ for the companion.

(4) The companion is located in the Hertzsprung gap on the HR diagram and far away from single-stellar isochrone for its age. The progenitor and the companion may have very similar initial masses and binary mass transfer took place when the companion had already evolved off the MS; alternatively, the companion may still be a MS star but temporarily out of thermal equilibrium due to very recent binary interaction.

(5) SN~2019yvr exhibits signatures of interaction with CSM ejected shortly before core collapse. Our progenitor constraint supports enhanced pre-SN mass loss as an important process for hydrogen-poor stars at the lower-mass end of core-collapse SN progenitors.

(5) SN~2019yvr may be the first Type~Ib/c SN with a direct detection of its binary companion. Late-time observations, when the SN has faded, will finally confirm the nature of the pre-explosion source. The source would (almost) completely disappear if the pre-explosion brightness were dominated by the progenitor star itself; alternatively, it would only become mildly fainter if the brightness were dominated by the binary companion as suggested in this work.

\section*{Acknowledgements}

We thank the anonymous referee for providing very helpful comments to our paper. N-CS's research is funded by the Science and Technology Facilities Council through grant ST/V000853/1. EZ acknowledges support by the Swiss National Science Foundation Professorship grant (project number PP00P2 176868; PI Tassos Fragos). This paper is based on observations made with the NASA/ESA Hubble Space Telescope and with the Very Large Telescope at the European Southern Observatory. This paper has also used the light curve of SN~2019yvr published in \citet{K21}.

\section*{Data availability}

Data used in this work are all publicly available from the ESO data archive (\url{http://archive.eso.org}), the Mikulski Archive for Space Telescope (\url{https://archive.stsci.edu}), and the paper by \citet{K21}.

\bsp	
\label{lastpage}
\end{document}